\documentclass[twocolumn,twocolappendix,times,tighten,trackchanges]{aastex631}
\usepackage{natbib}
\usepackage{xspace}
\usepackage{amsmath}
\usepackage{url}
\usepackage{appendix}

\newcommand{\uat}[2]{\href{http://astrothesaurus.org/uat/#2}{#1 (#2)}}

\newcommand{\lamlam}{$\lambda\lambda$}
\newcommand{\HII}{\mbox{H\,{\sc ii}}}

\newcommand{\CI}{\mbox{C\,{\sc i}}}
\newcommand{\CII}{\mbox{C\,{\sc ii}}}

\newcommand{\OII}{\mbox{O\,{\sc ii}}}
\newcommand{\OIII}{\mbox{O\,{\sc iii}}}
\newcommand{\NII}{\mbox{N\,{\sc ii}}}

\newcommand{\Hb}{\ensuremath{\mathrm{H\beta}}}
\newcommand{\Hgamma}{\ensuremath{\mathrm{H\gamma}}}
\newcommand{\Hdelta}{\ensuremath{\mathrm{H\delta}}}
\newcommand{\Lya}{\ensuremath{\mathrm{Ly\alpha}}}

\newcommand{\SFR}{\ensuremath{\mathrm{SFR}}}

\newcommand{\Zgas}{\ensuremath{Z_\text{gas}}}
\newcommand{\Zstar}{\ensuremath{Z_\text{star}}}
\newcommand{\Zsun}{\ensuremath{Z_\text{\(\odot\)}}}

\newcommand{\Umin}{\ensuremath{\mathcal{U}_\mathrm{min}}}

\newcommand{\kms}{\ensuremath{\mathrm{km\ s^{-1}}}}
\newcommand{\Mo}{\ensuremath{\mathrm{M_\sun}}}
\newcommand{\Msun}{\ensuremath{\mathrm{M_\sun}}}
\newcommand{\Moyr}{\ensuremath{\mathrm{M_\sun\ yr^{-1}}}}

\newcommand{\Moyrkpct}{\ensuremath{\mathrm{M_\sun\ yr^{-1}\ kpc^{-2}}}}
\newcommand{\Mopcpc}{\ensuremath{\mathrm{M_\sun\ pc^{-2}}}}

\newcommand{\ergscm}{\ensuremath{\mathrm{erg\ s^{-1}\ cm^{-2}}}}

\newcommand{\um}{$\mathrm{\mu m}$}

\newcommand{\nJy}{$\mathrm{nJy}$}

\newcommand{\uJybeam}{$\mathrm{\mu Jy\ beam^{-1}}$}
\newcommand{\Jybeamkms}{$\mathrm{Jy\ beam^{-1} km\ s^{-1}}$}

\newcommand{\cmmm}{\ensuremath{\mathrm{cm^{-3}}}}

\newcommand{\OH}{\ensuremath{\text{O}/\text{H}}}

\newcommand{\SN}{\ensuremath{ \text{S}/\text{N}}}

\newcommand{\betaUV}{\ensuremath{\beta_\text{UV}}}
\newcommand{\betaUVO}{\ensuremath{\beta_\text{UV,0}}}

\newcommand{\nn}{\mbox{--}}

\shorttitle{RIOJA.\@ Big Three Dragons}
\shortauthors{Sugahara et al.}

\begin{document}

\title{\large RIOJA.\@ Complex Dusty Starbursts in a Major Merger B14-65666 at z=7.15}

\author[0000-0001-6958-7856]{Yuma Sugahara}
\affil{National Astronomical Observatory of Japan, 2-21-1 Osawa, Mitaka, Tokyo 181-8588, Japan}
\affil{Waseda Research Institute for Science and Engineering, Faculty of Science and Engineering, Waseda University, 3-4-1 Okubo, Shinjuku, Tokyo 169-8555, Japan}
\affil{Department of Pure and Applied Physics, School of Advanced Science and Engineering, Faculty of Science and Engineering, Waseda University, 3-4-1 Okubo, Shinjuku, Tokyo 169-8555, Japan}
\email{sugayu@aoni.waseda.jp}
\author[0000-0002-7093-1877]{Javier \'Alvarez-M\'arquez}
\affil{Centro de Astrobiolog\'{\i}a (CAB), CSIC-INTA, Ctra. de Ajalvir km 4, Torrej\'on de Ardoz, E-28850, Madrid, Spain}
\author[0000-0002-0898-4038]{Takuya Hashimoto}
\affil{Division of Physics, Faculty of Pure and Applied Sciences, University of Tsukuba,Tsukuba, Ibaraki 305-8571, Japan}
\affil{Tomonaga Center for the History of the Universe (TCHoU), Faculty of Pure and Applied Sciences, University of Tsukuba, Tsukuba, Ibaraki 305-8571, Japan}
\author[0000-0002-9090-4227]{Luis Colina}
\affil{Centro de Astrobiolog\'{\i}a (CAB), CSIC-INTA, Ctra. de Ajalvir km 4, Torrej\'on de Ardoz, E-28850, Madrid, Spain}
\author[0000-0002-7779-8677]{Akio K. Inoue}
\affil{Waseda Research Institute for Science and Engineering, Faculty of Science and Engineering, Waseda University, 3-4-1 Okubo, Shinjuku, Tokyo 169-8555, Japan}
\affil{Department of Pure and Applied Physics, School of Advanced Science and Engineering, Faculty of Science and Engineering, Waseda University, 3-4-1 Okubo, Shinjuku, Tokyo 169-8555, Japan}
\author[0000-0001-6820-0015]{Luca Costantin}
\affil{Centro de Astrobiolog\'{\i}a (CAB), CSIC-INTA, Ctra. de Ajalvir km 4, Torrej\'on de Ardoz, E-28850, Madrid, Spain}
\author[0000-0001-7440-8832]{Yoshinobu Fudamoto}
\affil{National Astronomical Observatory of Japan, 2-21-1 Osawa, Mitaka, Tokyo 181-8588, Japan}
\affil{Waseda Research Institute for Science and Engineering, Faculty of Science and Engineering, Waseda University, 3-4-1 Okubo, Shinjuku, Tokyo 169-8555, Japan}
\affil{Center for Frontier Science, Chiba University, 1-33 Yayoi-cho, Inage-ku, Chiba 263-8522, Japan}
\author[0000-0003-4985-0201]{Ken Mawatari}
\affil{Division of Physics, Faculty of Pure and Applied Sciences, University of Tsukuba,Tsukuba, Ibaraki 305-8571, Japan}
\affil{Tomonaga Center for the History of the Universe (TCHoU), Faculty of Pure and Applied Sciences, University of Tsukuba, Tsukuba, Ibaraki 305-8571, Japan}
\author[0000-0002-6510-5028]{Yi W. Ren}
\affil{Department of Pure and Applied Physics, School of Advanced Science and Engineering, Faculty of Science and Engineering, Waseda University, 3-4-1 Okubo, Shinjuku, Tokyo 169-8555, Japan}
\author[0000-0001-7997-1640]{Santiago Arribas}
\affil{Centro de Astrobiolog\'{\i}a (CAB), CSIC-INTA, Ctra. de Ajalvir km 4, Torrej\'on de Ardoz, E-28850, Madrid, Spain}
\author[0000-0002-5268-2221]{Tom J. L. C. Bakx}
\affil{Department of Space, Earth and Environment, Chalmers University of Technology, Onsala Space Observatory, SE-439 92 Onsala, Sweden}
% \affil{Department of Physics, Graduate School of Science, Nagoya University, Nagoya 464-8602, Japan}
% \affil{National Astronomical Observatory of Japan, 2-21-1, Osawa, Mitaka, Tokyo, Japan}
\author[0009-0005-5448-5239]{Carmen Blanco-Prieto}
\affil{Centro de Astrobiolog\'{\i}a (CAB), CSIC-INTA, Ctra. de Ajalvir km 4, Torrej\'on de Ardoz, E-28850, Madrid, Spain}
\author[0000-0002-8680-248X]{Daniel Ceverino}
\affil{Universidad Autonoma de Madrid, Ciudad Universitaria de Cantoblanco, E-28049 Madrid, Spain}
\affil{CIAFF, Facultad de Ciencias, Universidad Autonoma de Madrid, E-28049 Madrid, Spain}
\author[0000-0003-2119-277X]{Alejandro Crespo G\'omez}
\affil{Centro de Astrobiolog\'{\i}a (CAB), CSIC-INTA, Ctra. de Ajalvir km 4, Torrej\'on de Ardoz, E-28850, Madrid, Spain}
\author[0000-0001-8083-5814]{Masato Hagimoto}
\affil{Department of Physics, Graduate School of Science, Nagoya University, Nagoya, Aichi 464-8602, Japan}
\author{Takeshi Hashigaya}
\affil{Department of Astronomy, Kyoto University Sakyo-ku, Kyoto 606-8502, Japan}
\author[0000-0001-8442-1846]{Rui Marques-Chaves}
\affil{Geneva Observatory, Department of Astronomy, University of Geneva, Chemin Pegasi 51, CH-1290 Versoix, Switzerland}
\author[0000-0003-3278-2484]{Hiroshi Matsuo}
\affil{National Astronomical Observatory of Japan, 2-21-1 Osawa, Mitaka, Tokyo 181-8588, Japan}
\affil{The Graduate University for Advanced Studies (SOKENDAI), 2-21-1 Osawa, Mitaka, Tokyo 181-8588, Japan}
\author[0000-0002-0984-7713]{Yurina Nakazato}
\affil{Department of Physics, The University of Tokyo, 7-3-1 Hongo, Bunkyo, Tokyo 113-0033, Japan}
\author[0000-0002-4005-9619]{Miguel Pereira-Santaella}
\affil{Instituto de F\'isica Fundamental (IFF), CSIC, Serrano 123, E-28006, Madrid, Spain}
\author[0000-0003-4807-8117]{Yoichi Tamura}
\affil{Department of Physics, Graduate School of Science, Nagoya University, Nagoya, Aichi 464-8602, Japan}
\author{Mitsutaka Usui}
\affil{Division of Physics, Faculty of Pure and Applied Sciences, University of Tsukuba,Tsukuba, Ibaraki 305-8571, Japan}
\author[0000-0001-7925-238X]{Naoki Yoshida}
\affil{Department of Physics, The University of Tokyo, 7-3-1 Hongo, Bunkyo, Tokyo 113-0033, Japan}
\affil{Kavli Institute for the Physics and Mathematics of the Universe (WPI), UT Institute for Advanced Study, The University of Tokyo, Kashiwa, Chiba 277-8583, Japan}
\affil{Research Center for the Early Universe, School of Science, The University of Tokyo, 7-3-1 Hongo, Bunkyo, Tokyo 113-0033, Japan}
% \author{RIOJA team}

% \author[0000-0002-7779-8677]{Akio K. Inoue}
% \affil{Waseda Research Institute for Science and Engineering, Faculty of Science and Engineering, Waseda University, 3-4-1, Okubo, Shinjuku, Tokyo 169-8555, Japan}
% \affil{Department of Physics, School of Advanced Science and Engineering, Faculty of Science and Engineering, Waseda University, 3-4-1, Okubo, Shinjuku, Tokyo 169-8555, Japan}
% \author[0000-0001-7440-8832]{Yoshinobu Fudamoto}
% \affil{National Astronomical Observatory of Japan, 2-21-1 Osawa, Mitaka, Tokyo 181-8588, Japan}
% \affil{Waseda Research Institute for Science and Engineering, Faculty of Science and Engineering, Waseda University, 3-4-1, Okubo, Shinjuku, Tokyo 169-8555, Japan}
% \author[0000-0002-0898-4038]{Takuya Hashimoto}
% \affil{Tomonaga Center for the History of the Universe (TCHoU), Faculty of Pure and Applied Sciences, University of Tsukuba, Tsukuba, Ibaraki 305-8571, Japan}

%%% Local Variables:
%%% mode: japanese-latex
%%% TeX-master: "note"
%%% End:

\begin{abstract} % 250 words [tex-count-words] % 1920 characters in arXiv
We present JWST NIRCam imaging of B14-65666 (``Big Three Dragons''), a bright Lyman-break galaxy system (\(M_\text{UV}=-22.5\)~mag) at \(z=7.15\).
The high angular resolution of NIRCam reveals the complex morphology of two galaxy components:
galaxy E has a compact core (E-core), surrounded by diffuse, extended, rest-frame optical emission, which is likely to be tidal tails;
and galaxy W has a clumpy and elongated morphology with a blue UV slope (\(\betaUV=-2.2\pm0.1\)).
The flux excess, F356W\(-\)F444W, peaks at the E-core (\(1.05^{+0.08}_{-0.09}\)~mag), tracing the presence of strong [\OIII]\lamlam4960,5008~\AA\ emission.
ALMA archival data show that the bluer galaxy W is brighter in dust continua than the redder galaxy E, while the tails are bright in [\OIII]~88~\um.
The UV/optical and sub-mm SED fitting confirms that B14-65666 is a major merger in a starburst phase as derived from the stellar mass ratio (3:1 to 2:1) and the star-formation rate, \(\simeq1\)~dex higher than the star-formation main sequence at the same redshift.
The galaxy E is a dusty (\(A_\text{V}=1.2\pm0.1\)~mag) starburst with a possible high dust temperature (\(\ge63\nn68\)~K).
The galaxy W would have a low dust temperature (\(\le27\nn33\)~K) or patchy stellar-and-dust geometry, as suggested from the infrared excess (IRX) and \betaUV\ diagram.
The high optical-to-FIR [\OIII] line ratio of the E-core shows its lower gas-phase metallicity (\(\simeq0.2\nn0.4\)~\Zsun) than the galaxy W.
These results agree with a scenario where major mergers disturb morphology and induce nuclear dusty starbursts triggered by less-enriched inflows.
B14-65666 shows a picture of complex stellar buildup processes during major mergers in the epoch of reionization.
\end{abstract}


\section{Introduction}\label{sec:introduction}

Galaxy mergers are a fundamental process to build up the stellar content through galaxy evolution.
Gravitational perturbations by galaxy interactions can strongly disturb the morpho-kinematics, which is characterized by tidal tails and large velocity dispersion.
It is theoretically predicted that such perturbation promotes gas accretion from outskirts into the central parts of galaxies that can trigger nuclear star formation \citep{Mihos.J:1994a, Mihos.J:1996a, Barnes.J:1996a}.
In observations of local galaxies, the star-formation rate (\SFR) of interacting galaxies increases as projected pair separation decreases to within several tens kpc \citep[e.g.,][]{Barton.E:2000a}.
This \SFR\ enhancement is particularly notable in collisions of galaxies with similar stellar masses \citep[i.e., ``major mergers''; e.g.,][]{Ellison.S:2008a}.
The gas accretion associated with mergers can also induce an observed reduction in the gas-phase metallicity at a fixed mass or luminosity \citep[e.g.,][]{Kewley.L:2006a, Ellison.S:2008a}.
Even after the coalescence phase, the major mergers are expected to cause energetic events in the subsequent evolution of the merged galaxies, such as luminous infrared starbursts, rapid black hole growths, and the appearance of quasars \citep{Hopkins:2008a}.

At high redshifts, major mergers are predicted to play a more important role in the evolutionary path because the merger rate theoretically increases with increasing redshift by \(\propto (1 + z)^{2\nn3}\) \citep[e.g.,][]{Fakhouri.O:2008a, Rodriguez-Gomez.V:2015a}.
This increase is consistent with observed major merger fractions at \(z\sim0\nn5\) \citep[e.g.,][]{Man.A:2016a, Ventou.E:2017a} when the redshift-dependent merger timescale is taken into account \citep{Qu.Y:2017a, Snyder.G:2017a, Romano.M:2021a}.
Recently, rest-frame ultraviolet (UV) images \citep{Ribeiro.B:2017a, Duncan.K:2019a, Shibuya.T:2022a} and [\CII] 158 \um\ data cubes \citep{Romano.M:2021a} were able to explore the merger fractions up to \(z \sim 5\nn7\).
Although the measured fractions are yet to be converged and depend strongly on galaxy masses, the high merger fraction, \(20\nn80\)\% at \(z\sim5\nn7\), supports a large number of galaxies that experience major mergers in the early Universe.
In terms of the stellar mass assembly, contributions of major mergers are suggested to be comparable to in-situ star formation at \(z > 3\) \citep{Duncan.K:2019a, Romano.M:2021a, Duan.Q:2024a}.
While cosmological zoom-in simulations also support the SFR enhancement by gas-rich mergers at \(z > 5\) \citep{Ceverino.D:2018a}, the effect of mergers on star formation appears to reduce strongly from the local Universe to \(z\sim1\nn3\) \citep[e.g.,][]{Silva.A:2018a, Shah.E:2022a}.

High angular resolution is key to pushing up the redshift frontier of merger studies.
Merging systems at \(z\gtrsim5\) are reported from observations of Hubble Space Telescope (HST), James Webb Space Telescope \citep[JWST,][]{Gardner.J:2023a}, and Atacama Large Millimeter/sub-millimeter Array (ALMA), which have both high angular resolutions and highest sensitivities that can resolve high-redshift galaxies.
For example, dusty systems at \(z\sim4\nn7\) corresponding to hyper/ultra luminous infrared galaxies (ULIRGs) sometimes show signatures of galaxy mergers \citep[e.g.][]{Oteo.I:2016a, Marrone.D:2018a, Hygate.A:2023a, Bik.A:2024a}.
Triple mergers have also been observed, such as a dust-free, luminous \Lya\ emitter at \(z = 6.6\) \citep[``Himiko'',][]{Ouchi:2013a} and the [\CII]-bright galaxy at \(z = 4.6\) \citep{Jones.G:2020a}.
JWST has recently updated the most distant major merger by spectroscopically confirming a triply-lensed galaxy at \(z = 10.17\) that has clumps with different dust attenuation \citep{Hsiao.T:2023a, Hsiao.T:2023b}.
For a statistical study, JWST has shown that galaxy interactions significantly contribute to the bursty star-formation and strong emission lines in low-mass galaxies at \(z\sim5\nn6\) \citep{Asada.Y:2024a}.
However, few merging galaxies are observed and detected with multiple bands in both UV and far-infrared (FIR), limiting our ability to interpret the complete properties of distant merging systems.

Multi-wavelength, high angular-resolution observations are essential to examine the complex properties of stars, gas, and dust through UV-to-FIR spectral energy distributions (SEDs) of resolved galaxy components.
Red colors of high-redshift galaxies are generally tied with dust attenuation similar to local star-forming galaxies \citep{Bowler.R:2022a}, but some galaxies exhibit spatially patchy geometry of UV, line, and dust emission \citep[e.g.,][]{Tamura.Y:2023b, Colina.L:2023a}.
This complexity as well as undetermined dust attenuation curves at high redshift strengthens the importance of multi-wavelength observations.
Under these circumstances, combinations of JWST and ALMA observations offer valuable constraints on SEDs of spatially resolved galaxies.
JWST near-infrared observations uncover resolved stellar and gaseous properties of galaxies even at \(z > 6\) \citep{Gimenez-Arteaga.C:2023a, Abdurrouf:2023a}.
Dust detections with (multiple) ALMA bands help break the degeneracy between the star-formation age and dust attenuation by constraining the dust temperature and FIR \SFR\ \citep[e.g.,][]{Bakx.T:2021a, Alvarez-Marquez.J:2023a, Li.J:2024a, Crespo-Gomez.A:2024a}.
Moreover, these combinations allow us to estimate nebular properties of the interstellar media with optical-to-FIR emission line ratios \citep{Fujimoto.S:2024b, Stiavelli.M:2023a}.
Thus, the synergy between JWST and ALMA effectively works in revealing the complex properties of merging galaxies during the epoch of reionization (EoR).

This paper presents new observations of a bright merging galaxy system at \(z = 7.15\), B14-65666, also known as the Big Three Dragons \citep{Hashimoto.T:2019a}, with the Near Infrared Camera \citep[NIRCam;][]{Rieke.M:2023a} on board JWST\@.
These observations were conducted as part of the RIOJA project: ``the Reionization and the ISM/Stellar Origins with JWST and ALMA'' \citep[JWST GO1 PID1840; PIs: J. \'{A}lvarez-M\'{a}rquez and T. Hashimoto;][]{Alvarez-Marquez.J:2021b, Hashimoto.T:2023c}.
Combining ALMA ancillary data with the NIRCam images, we investigate stellar, gaseous, and dusty properties in morphology and star formation of B14-65666.
This paper is organized as follows.
Section~\ref{sec:observational-data} describes NIRCam and ancillary data, and reduction processes.
Section~\ref{sec:analyses} explains analysis tools including image analyses, SED fitting, and photoionization models.
Section~\ref{sec:results} presents results about morphology, comparison with ALMA data, and SED fitting.
Section~\ref{sec:discussion} discusses relations between UV and FIR photometry, a metallicity distribution, and our interpretations about B14-65666.
Finally, Section~\ref{sec:summary} summarizes our conclusions.
Throughout this paper, we use the AB magnitude system \citep{Oke.J:1983a} and the flat $\Lambda$CDM cosmology (\(\Omega_\mathrm{M} = 0.310\) and \(h = H_0/[100\ \mathrm{km\ s^{-1}\ Mpc^{-1}}] = 0.677 \); \citealt{Planck-Collaboration:2020a}).
At \(z = 7.15\), \(1\) arcsec corresponds to \(5.27\) kpc.

\section{Observations}\label{sec:observations}

\subsection{Summary of Previous Observations of B14-65666}\label{sec:targetb14-65666}

B14-65666 was first identified as a Lyman-break galaxy from the UltraVISTA survey \citep{Bowler.R:2012a, Bowler.R:2014a}.
Its UV absolute magnitude of \(M_\text{UV} = -22.5\)~mag is four times brighter than the characteristic value at \(z = 7\) \citep[\(M^{*}_\text{UV} \simeq -21.0\)~mag; e.g.,][]{Finkelstein.S:2015b, Bouwens.R:2021b}.
Multi-wavelength follow-up observations were conducted to detect the \Lya\ emission line \citep{Furusawa.H:2016a}, the UV continuum with high-spatial resolution \citep{Bowler.R:2017a}, the [\OIII] 88 \um\ and [\CII] 158 \um\ emission lines \citep{Hashimoto.T:2019a}, and dust continua at 88, 122, and 158 \um\ \citep{Bowler.R:2018a, Hashimoto.T:2019a, Sugahara.Y:2021a}.
From the dust continua, the infrared luminosity is estimated to be \(\log{L_\text{IR}/L_{\sun}} = 11.6^{+0.6}_{-0.3}\) at the dust emissivity of \(\beta_\text{d} = 2.0\) \citep{Sugahara.Y:2021a}.
The galaxy was also observed by ALMA Bands 3 and 7, but it remained undetected in CO(6-5), CO(7-6), [\CI](2-1) \citep{Hashimoto.T:2023b} and [\NII] 122 \um\ \citep{Sugahara.Y:2021a}.
The systemic redshift was determined from the [\OIII] and [\CII] emission lines to be \(z = 7.1520\pm0.0003\) \citep{Hashimoto.T:2019a}.
These observations do not provide any evidences towards the presence of an active galactic nucleus (AGN) in the B14-65666 system.
% These ancillary multi-wavelength data, particularly its rich ALMA data, make B14-65666 a valuable follow-up target.
We use our new NIRCam data and the ancillary data to investigate detailed merger properties during EoR.

\subsection{JWST and ALMA Observations and Data Reductions}\label{sec:observational-data}

The NIRCam images of B14-65666 were taken on Dec.~20, 2022 with a total exposure time of \(644\) s in F200W and F277W, \(730\) s in F115W and F356W, and \(1675\) s in F150W and F444W.
The F115W, F200W, F277W, and F356W observations adopted the BRIGHT1 readout pattern with eight or nine groups and one integration; the F150W and F444W observations adopted the SHALLOW4 with eight groups one integration.
All the exposures were four times dithered using the INTRAMODULEBOX pattern.
In the RIOJA project, the NIRCam filters were selected such as to trace the continuum emission over the entire \(1\)--\(5\) \um\ wavelength range with mostly avoiding the contamination by the main optical emission lines.
Given the spectroscopic redshift of \(z = 7.1520\), the short wavelength channels of F115W, F150W, and F200W correspond to the rest-frame ultraviolet (\(1400\nn2500\) \AA).
The F277W and F356W bands cover the stellar continua on both sides of the Balmer break without any contamination of strong optical emission lines.
The F444W band includes the \Hb\ and [\OIII]\lamlam4960,5008 \AA\ emission lines as well as underlying stellar continuum.

The data were reduced using our custom pipeline based on the JWST calibration pipeline version 1.9.4 \citep{Bushouse.H:2023a} under CRDS context jwst\_1041.pmap.
Our additional reduction steps are:
(1) snowballs and wisps removal as described in \citet{Bagley.M:2023a} and (2) background homogenization (including \(1/f\)-noise subtraction) applied in \citet{Perez-Gonzalez.P:2023b}.
The pixel scales of the final products for all the filters are set to \(0.03\) arcsec\(/\)pixel.
The measured angular resolution and \(5\sigma\) depths of the NIRCam filters are listed in Table~\ref{tb:size}.
We note that although the HST image has a comparable depth to the NIRCam ones, we did not use it in this study to avoid a potential systematics among the telescope.

\begin{deluxetable}{lccc}
    \tablecaption{Summary of NIRCam and ALMA observations.\label{tb:size}}
    \tabletypesize{\small}
    \tablewidth{0pt}
    \tablecolumns{4}
    \tablehead{
      \colhead{Band} & \colhead{\(\lambda_\text{rest}\)} & \colhead{PSF FWHM} & \colhead{\(5\sigma\) Depth}\\
      \colhead{} & \colhead{} & \colhead{[arcsec]} & \colhead{}
    }
    \startdata
    NIRCam/F115W & \(1416\)~\AA & \(0.060\) & \(23.3\)~\nJy \\
    NIRCam/F150W & \(1841\)~\AA & \(0.067\) & \(7.7\)~\nJy \\
    NIRCam/F200W & \(2441\)~\AA & \(0.076\) & \(23.5\)~\nJy \\
    NIRCam/F277W & \(3418\)~\AA & \(0.13\) & \(20.4\)~\nJy \\
    NIRCam/F356W & \(4371\)~\AA & \(0.15\) & \(19.5\)~\nJy \\
    NIRCam/F444W & \(5424\)~\AA & \(0.16\) & \(15.4\)~\nJy \\
    ALMA/Band8 & \(88.84\)~\um & \(0.39\times0.36\) & \(141\)~\uJybeam \\
    % ALMA/[OIII]88 & \(88.35\)~\um & \(0.46\times0.41\) & \(0.000\)~\uJybeam \\
    ALMA/Band7 & \(122.6\)~\um & \(1.17\times0.98\) & \(45.5\)~\uJybeam \\
    ALMA/Band6 & \(158.7\)~\um & \(0.28\times0.20\) & \(42.3\)~\uJybeam \\
    \enddata
    \tablecomments{The rest-frame wavelengths, \(\lambda_\text{rest}\), are converted from the pivot wavelengths for NIRCam.
      The \(5\sigma\) depths are computed within \(2\times\)PSF-FWHM diameter apertures for NIRCam and with a peak rms for ALMA.}
\end{deluxetable}

%%% Local Variables:
%%% mode: latex
%%% TeX-master: "../ms"
%%% End:
 % tb:size
\vspace{-1\baselineskip}

The ALMA data used in this work were presented in previous studies.
ALMA Bands 6 and 8 data were taken in Cycles 4 (ID 2016.1.00954.S, PI:\@ A. K. Inoue) and 5 (ID 2017.1.00190.S, PI:\@ A. K. Inoue) to target the [\CII] 158 \um\ and [\OIII] 88 \um\ emission lines \citep{Hashimoto.T:2019a}.
ALMA Band 7 data were taken in Cycle 7 (ID:\@ 2019.1.01491.S, PI:\@ A. K. Inoue) to target the [\NII] 122 \um\ emission line \citep{Sugahara.Y:2021a}.
In all the observations, one of the four spectral windows was placed to observe the targeted lines while the others were for the underlying dust continua.
The raw data were reduced with standard scripts on Common Astronomy Software Applications \citep[][]{CASA-Team:2022a}.
[\OIII] and [\CII] emission-line maps (i.e., moment-0 maps) were created by integrating wavelength ranges of \(\simeq600\) \kms\ around the line centers after the continuum subtraction \citep{Hashimoto.T:2019a}.
Dust continuum maps were created using channels that are unaffected by the emission lines.
The synthesized beam FWHM is \(0\farcs29\times0\farcs23\), \(1\farcs15\times0\farcs97\), and \(0\farcs39\times0\farcs37\) in Bands 6, 7, and 8 observations with the natural weighting.
More information on the observations and data reduction is described in the references above.
We did not use ALMA Band 3 data \citep[ID: 2018.1.01673.S, PI: T. Hashimoto;][]{Hashimoto.T:2023b} in this study because neither continuum nor lines were detected and they do not give strong constraints on the FIR SED.
The beam sizes and \(5\sigma\) depths for the peak \SN\ of the ALMA observations are listed in Table~\ref{tb:size}.

Astrometry of the NIRCam images were corrected with stars in the field of view (FoV).
In the NIRCam FoV, there exist three stars listed on the Gaia DR3 main catalog \citep{Gaia-Collaboration:2016a, Gaia-Collaboration:2023a}.
We aligned the NIRCam images to the three Gaia stars, and then we corrected small offsets between the NIRCam images using faint stars.
The astrometric uncertainties of the NIRCam images are \( < 0\farcs015\).
The peak positional accuracy of ALMA data, which can be estimated from the beam size and peak \SN\footnote{Equation (10.7) in ALMA Cycle 10 Technical Handbook: \url{https://almascience.nao.ac.jp/documents-and-tools/cycle10/alma-technical-handbook}}, is \(\sim0\farcs03\) for [\CII] 158 \um\ and [\OIII] 88 \um\ line maps and \(\sim0\farcs06\) for dust continuum maps at \(158\) and \(88\) \um.
As these estimates for the ALMA data are based on an equation for compact sources, the accuracy may be worse for extended sources like B14-65666.

\section{Analyses}\label{sec:analyses}

\subsection{Image analyses}\label{sec:image-analyses}
When comparing different bands of the NIRCam images pixel by pixel, we matched the point spread functions (PSFs) of the images to that of the F444W image by kernel convolution.
We constructed effective PSFs following the \citet{Anderson.J:2000a} and \citet{Anderson.J:2016a} methods using \texttt{photutils} \citep{Bradley.L:2023a}, which construct empirical models by fitting pixel values in images of stars detected with \texttt{IRAFStarFinder}.
Then, we homogenized the PSFs of the NIRCam images using kernels created by \texttt{PyPHER} \citep{Boucaud.A:2016a}.
We have checked that the PSF homogenization works well using faint stars in FoVs.
Finally, the image pixels were resampled using \texttt{reproject} to match the F444W image pixels.
When comparing the NIRCam and ALMA [\OIII] images, the PSFs of the NIRCam images are homogenized to the synthesized clean beam (i.e., two-dimensional Gaussian) of the ALMA images following the same procedure.
The ALMA image pixels are resampled to match the F444W image pixels.

\begin{figure*}[t]
    \epsscale{1.2}
    \plotone{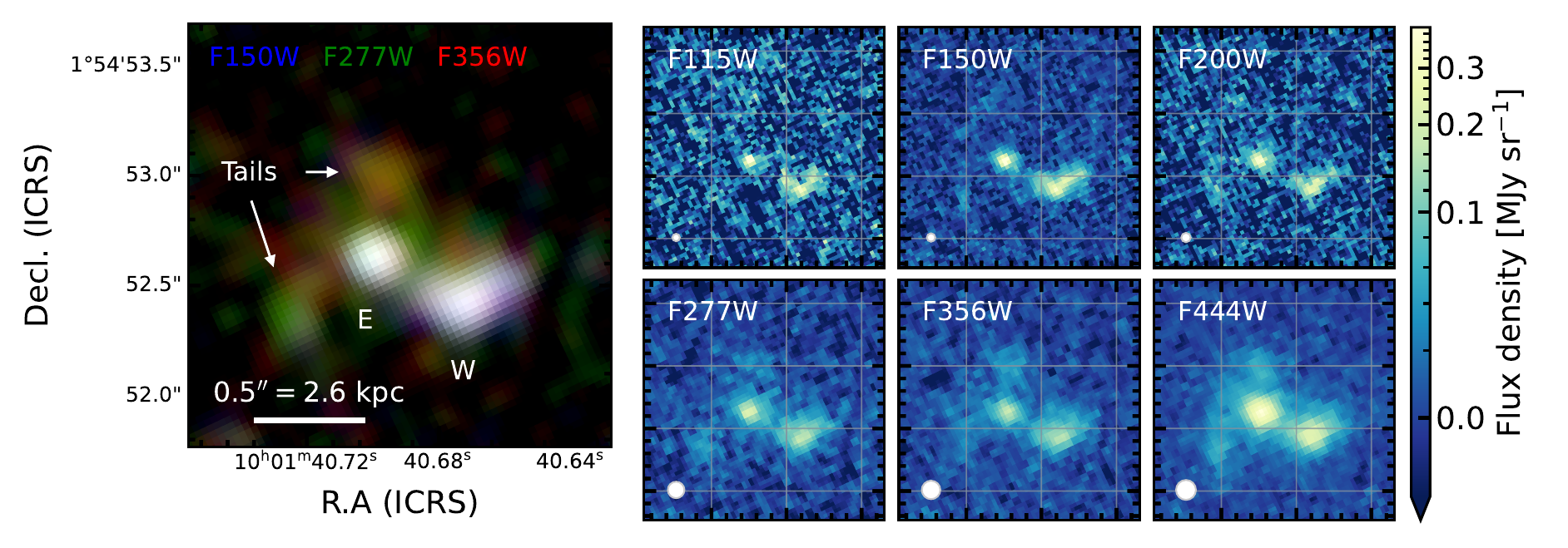}
    \caption{
      Left: Pseudo-RGB composite image of B14-65666, where F356W (red), F277W (green), and F150W (blue) bands reflect continuum light.
      The RGB images are convolved to match the spatial resolution of the F444W-band image.
      B14-65666 shows complex morphology including the elongated galaxy W and the galaxy E, which has a compact core being surrounded by the red tails.
      Right: NIRCam band stamps of B14-65666, which show the same sky area as the left panel; the grids intervals are \(0\farcs6\) in R.A. and \(0\farcs5\) in Decl.
      They are illustrated in the native PSFs, which are shown with the white circles at the bottom lefts.
    }\label{fig:nircam}
\end{figure*}

Photometry was performed with fixed apertures on the PSF-homogenized images and aperture correction was not applied.
We chose a circular aperture of \(1\farcs8\) in diameter to measure the total photometry of B14-65666, which sufficiently recover all the fluxes according to curve-of-growth plots of the flux density.
Because B14-65666 has a complex structure with several components and diffuse emission (see Section~\ref{sec:results} and Figure~\ref{fig:nircam}), smaller apertures are used to measure the photometry of these individual components (see Figure~\ref{fig:o3}).
The local sky background computed in an annulus from \(0\farcs95\) to \(1\farcs95\) were subtracted before the photometry measurements.
The photometric uncertainties were estimated from random aperture photometry.
We scattered the apertures over the images around B14-65666 where bright objects are masked and we computed the uncertainties as the standard deviation of the aperture-photometry distributions.
In addition, the absolute flux uncertainty was included in the photometric uncertainties as a conservative value of \(5\)\% \citep{Rigby.J:2023a}\footnote{See also ``NIRCam Imaging Calibration Status'' in JWST User Documentation for the current photometric calibration: \url{https://jwst-docs.stsci.edu}}.

\subsection{SED fitting}\label{sec:sed-fitting}

We performed SED fitting to the NIRCam photometry using Bayesian Analysis of Galaxies for Physical Inference and Parameter EStimation \citep[\texttt{Bagpipes},][]{Carnall.A:2018a}.
We tested SED fitting with and without ALMA fluxes to see changes in obtained best-fit parameters.
The ALMA [\OIII] 88 \um\ line and continuum fluxes were taken from \citet{Hashimoto.T:2019a}.
We note that these ALMA fluxes were measured by fitting two Gaussian components to the line and continuum maps, which differs from the aperture photometry for the NIRCam images.
The redshift is fixed at \(z = 7.1520\) in the fitting.

The fitting parameters are the star-formation age \(t_\text{SF}\), the stellar mass \(M_{*}\), the ionization parameter \(U\), the \textit{V}-band dust attenuation \(A_\text{V}\), the stellar metallicity \(Z_{*}\), and the mass of an old stellar component \(M_\text{old}\).
The star-formation history of B14-65666 is simply assumed to be constant for \(t_\text{SF}\) down to \(z = 7.1520\).
We also tested different star-formation histories, but the fitting results were insensitive to different star-formation histories (Appendix \ref{sec:A1}), because B14-65666 is dominated by a young bursty star formation of \(\sim10\) Myr timescale.
The gas-phase metallicity is assumed to be identical to the stellar metallicity.
The ionization parameter \(U\) controls nebular emission-line ratios, which is implemented using a photoionization code \textsc{Cloudy} \citep{Ferland:1998a, Ferland.G:2017a}.
The hydrogen density in the \textsc{Cloudy} models is fixed to be \(n_\text{H} = 100\) \cmmm\ in \texttt{Bagpipes}.
The dust attenuation curve follows the \citet{Calzetti:2000} law.
In addition, considering that B14-65666 is a major merger, stellar components older than recent starbursts should exist.
To constrain the mass of the old stellar components, we added another constant star formation lasting for \(500\) Myr, from \(600\) Myr ago to \(100\) Myr ago with respect to \(z = 7.15\), which does not strongly contribute to UV continuum and emission-line fluxes.
This selected star-formation timescale is arbitrary, but changes of this timescale do not affect our conclusion.
The stellar metallicity of the old components is fixed to \(\Zstar = 0.5\) \Zsun\ \citep[assuming that \(Z = 0.02\) for solar metallicity;][]{Asplund.M:2021a} to reduce the number of free parameters, but we checked that results do not vary significantly with the assumed \Zstar\ value.
The best-fit parameters are computed as the median of the posterior distributions.
The \(1\sigma\) fitting uncertainties are evaluated from the \(16\) and \(84\) percentiles of the posterior distributions.

For FIR SED, we added fitting parameters of the minimum stellar intensity \Umin\ and the fraction \(\gamma\) of the dust mass heated by \(\mathcal{U} > \Umin\).
\texttt{Bagpipes} has adopted the \citet{Draine.B:2007a} dust emission models.
Instead of the originally adopted ones, we used the updated version \citep{Draine.B:2014a} implemented in the \texttt{CIGALE} SED fitting code \citep{Boquien.M:2019a}, which allows up to \(\Umin = 50\).
The model parameters other than \Umin\ and \(\gamma\) were fixed: the power-law index of the stellar intensity \(\alpha_\mathcal{U} = 2\), the maximum stellar intensity \(\mathcal{U}_\text{max} = 10^7\), and the PAH mass fraction \(q_\text{PAH} = 2.5\)\%.
The former two fixed parameters are appropriate to reproduce high-redshift galaxies \citep{Burgarella.D:2022b} as well as nearby galaxies \citep{Draine.B:2014a} and the latter \(q_\text{PAH}\) was fixed because our observations are insensitive to mid-IR wavelengths.
We note that while we added two fitting parameters for the FIR SED, dust continua are constrained at only \(2\nn3\) bands.
Thus, adding ALMA dust continua in the SED fitting is not quite effective to constrain the stellar properties better (as shown in Section~\ref{sec:result-sed-fitting}), but it helps estimate the dust temperature necessary for the energy balance between dust attenuation and emission.

For input stellar models, the Binary Population and Spectral Synthesis code \citep[\texttt{BPASS},][]{Eldridge.J:2017a} version 2.3 \citep{Byrne.C:2022a} was used.
\texttt{BPASS} includes effects of binary populations, which are important to reproduce strong [\OIII] emission lines observed at high redshift \citep{Sugahara.Y:2022a}.
Specifically, the version 2.3 implements high \(\alpha/\text{Fe}\) cases, reflecting abundance patterns found at \(z \gtrsim 2\) dominated by core-collapse supernovae \citep{Steidel:2014,Steidel:2016}.
This enhancement of \(\alpha\) elements is also key to reproducing [\OIII] emission line strengths and hard UV spectral slopes at high redshift \citep[e.g.,][]{Shapley.A:2019a, Sugahara.Y:2022a}.
We used the case of \(\varDelta \log(\alpha/\text{Fe}) = +0.6\), where the oxygen abundance is almost the solar value \(12 + \log\OH = 8.69\) \citep{Asplund.M:2021a} at \(Z = 0.02\) \citep{Byrne.C:2022a}.
The adopted initial mass function (IMF) is a \citet{Kroupa:2001} IMF in the mass range from \(0.1\) to \(300\) \Mo.
Their choice of the upper mass limit leads to a 45\% higher ionizing flux and stronger hydrogen recombination lines than a choice of an upper limit of \(100\) \Mo\ \citep{Stanway.E:2016a}.
The nebular-continuum and emission-line models are generated based on these stellar models.

\subsection{Photoionization models with \textsc{Cloudy}}\label{sec:phot-models}
In our emission-line modeling, we used a photoionization code \textsc{Cloudy} version 17.02 \citep{Ferland:1998a, Ferland.G:2017a}.
We note that our setup is more flexible than the one used in the \texttt{Bagpipes} SED fitting, by including an additional free parameter of \(n_\text{H}\).
We assumed a plane-parallel geometry under the constant pressure and elemental and dust-grain abundances in the \HII\ regions that are stored in \textsc{Cloudy}.
For the helium abundance, we simply added stellar yields to Big Bang nucleosynthesis as a function of the gas-phase metallicity \citep{Groves.B:2004a}.
The input ionizing spectra were constructed from the \texttt{BPASS} version 2.3 and the stellar metallicity is assumed to be identical to the gas-phase metallicity.
The star-formation history is set to be a \(10\)-Myr constant star formation.
Adopted variable parameters are the ionization parameter \(U\) and hydrogen density \(n_\text{H}\) at the illuminated surface, and the gas-phase metallicity \Zgas.
The calculations stop at the edge of the \HII\ region where the electron fraction is \(10^{-2}\).
The output emission-line fluxes are normalized by the \Hb\ line fluxes.

\begin{figure*}[t]
    \epsscale{1.2}
    \plotone{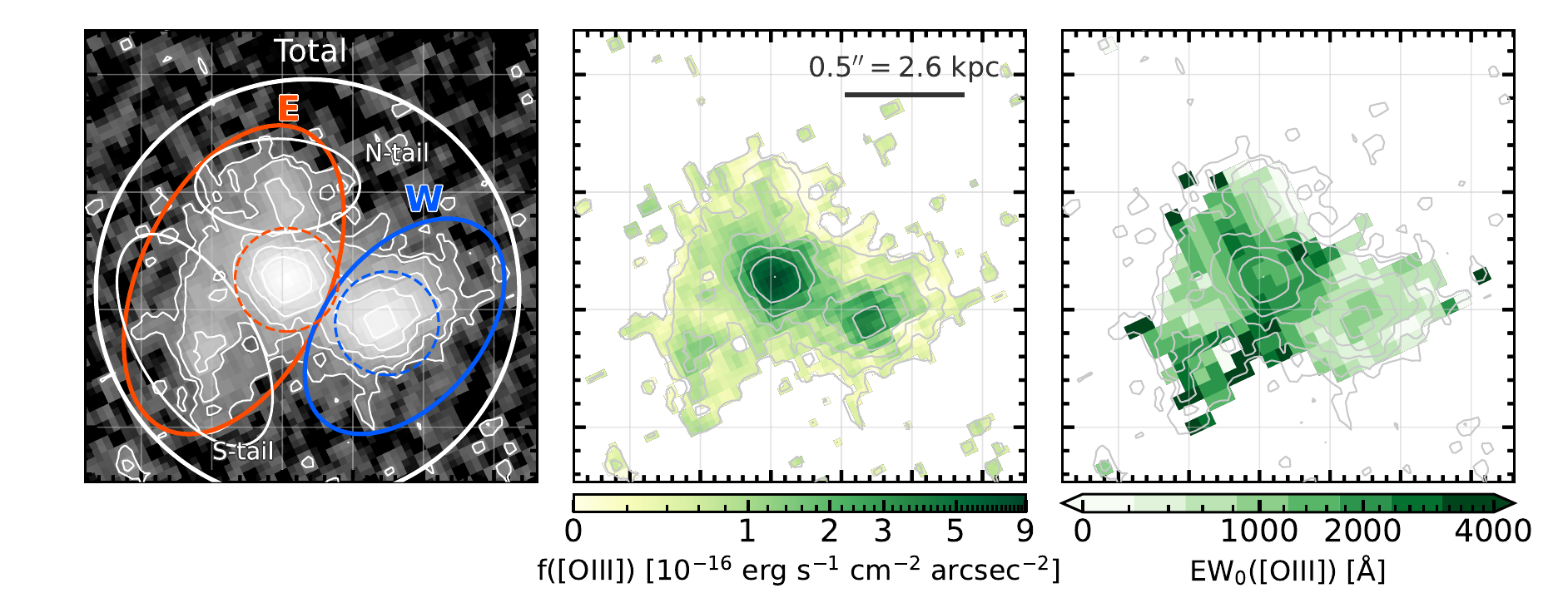}
    \caption{Left: Apertures used for photometry of different components in B14-65666, on the F444W image.
      The total aperture (white) contains flux from the entire system, and the galaxies E and W are represented with the red and blue circles, respectively.
      The cores of the galaxies are depicted with the dashed circles.
      The thin white circles show the tails (N-tail and S-tail).
      The background contours are at the \(2^n\sigma\) significance levels (\(n = 1, 2, 3, \ldots\)).
      Middle: [\OIII]\lamlam4960,5008 flux map estimated from the F356W and F444W images (see text for details).
      The flux map is drawn in pixels with \(\SN > 2\) in the F444W image.
      The gray contours are the F444W flux map, which is the same as in the left panel.
      Right: Rest-frame [\OIII] equivalent width map, which is re-binned to improve \SN\ per pixel.
      The equivalent widths are estimated by dividing the flux map by the stellar continuum inferred from F356W.
      The [\OIII] fluxes reach the peaks at the cores of the galaxies E and W.
      The rest-equivalent width values are \(\text{EW}_{0}([\OIII]) = 2050\pm290\) and \(740\pm170\) \AA\ for the E-core and W-core, respectively, which implies the presence of the nuclear starbursts induced by the major merger.
      The equivalent widths in the tails are not significant in the \(3\sigma\) level (Table~\ref{tb:properties}).
    }\label{fig:o3}
\end{figure*}

\section{Results}\label{sec:results}

\subsection{Morphology, Colors, and Optical [\OIII] Line Map}\label{sec:result-morphology}

B14-65666 consists of two bright components: galaxies E (east) and W (west), which were referred to as the clumps A and B in \citet{Hashimoto.T:2019a}, respectively.
\citet{Hashimoto.T:2019a} reported that these two galaxies have similar UV, IR, [\OIII] 88 \um, and [\CII] 158 \um\ luminosities within a factor of two and their line-of-sight velocity separation is \(\simeq200\) \kms.
For these reasons, B14-65666 is thought to be a major merger at \(z = 7.1520\).
The NIRCam observations provide us with morphological evidences for the scenario of the major merger.
Figure~\ref{fig:nircam} shows the NIRCam-band color image and the individual images in the six NIRCam filters of the two galaxies of B14-65666.
The two galaxies are apparently separated in the rest-frame optical wavelengths, ruling out a scenario that the two bright components are star-forming regions in a rotating stellar disk.

The NIRCam imaging clearly illustrates that the galaxies E and W have different morphology.
The galaxy E has a bright compact core.
This compact core is unresolved even in the F115W image, which has the highest angular resolution in the images used in this work.
Under an assumption that the core of the galaxy E has an exponential surface brightness profile, the deconvolved effective radius in F115W is \(r_\text{e} < 0.016\) arcsec (\(85 \) pc) at the \(2\sigma\) significance limit.
This core effective radius is smaller than the average effective radii (\(0.7\nn1\) kpc at rest-frame optical) of observed and simulated galaxies at \(M_{*}\sim10^9\) \Mo\ at \(z = 7\) \citep{Costantin.L:2023a, Ormerod.K:2024a}.

In contrast, the galaxy W is elongated from southeast to northwest.
This feature is seen in all the NIRCam images and its length is \(\simeq 0.3\) arcsec (\(1.5\) kpc).
Such elongated morphology look like tidal tails and it is often observed under tidal effect like major mergers in the local Universe \citep[e.g.,][]{Borne.K:2000a, Garcia-Marin.M:2009a}.
In addition, the F115W and F150W images suggest a clumpy internal structure of the galaxy W, which may consist of three clumps.
These compactness and clumpiness of B14-65666 agree with rest-frame UV clumps in bright high-redshift galaxies revealed by recent JWST observations \citep[e.g.,][]{Treu.T:2023a, Trussler.J:2023a, Marques-Chaves.R:2024a} and numerical simulations \citep[e.g.,][]{Mandelker.N:2017a, Ceverino.D:2023a, Nakazato.Y:2024a}.

In addition to the compact and clumpy structure, the NIRCam images have identified some extended emission around the galaxy E.
This emission is clearly seen in the F444W image and consists of two diffuse sub-components north and south-east of the galaxy E (Figure~\ref{fig:nircam}).
Actually, the spatially extended components were already suggested in the HST F140W image \citep{Bowler.R:2017a} and the [\OIII] 88 \um\ line map \citep{Hashimoto.T:2019a}, but they were not explicitly reported in the literature due to low \SN\@.
In the NIRCam imaging, the diffuse components are confirmed at \(\SN > 6\) in the F277W, F356W, and F444W bands and they are also suggested at lower significance levels (\(\SN \simeq 1\nn2\)) in the F150W and F200W images.
The F444W bandpass includes \Hb\ and [\OIII]\lamlam4960,5008 emission lines.
The F277W bandpass samples shorter wavelengths than the Balmer break, at which the stellar continuum by old stars becomes faint.
Therefore, these multiple band detections imply that the diffuse sub-components are bright in young stellar continuum or nebular emission (e.g., the [\OIII] and [\OII]\lamlam3727,3730 \AA\ lines and nebular continuum).
Given that B14-65666 is a major merging system, we assumed that they are likely to be tidal tails stripped from the galaxies \citep[e.g.,][]{Garcia-Marin.M:2009a}.

\begin{figure*}[t]
    \epsscale{1.0}
    \plotone{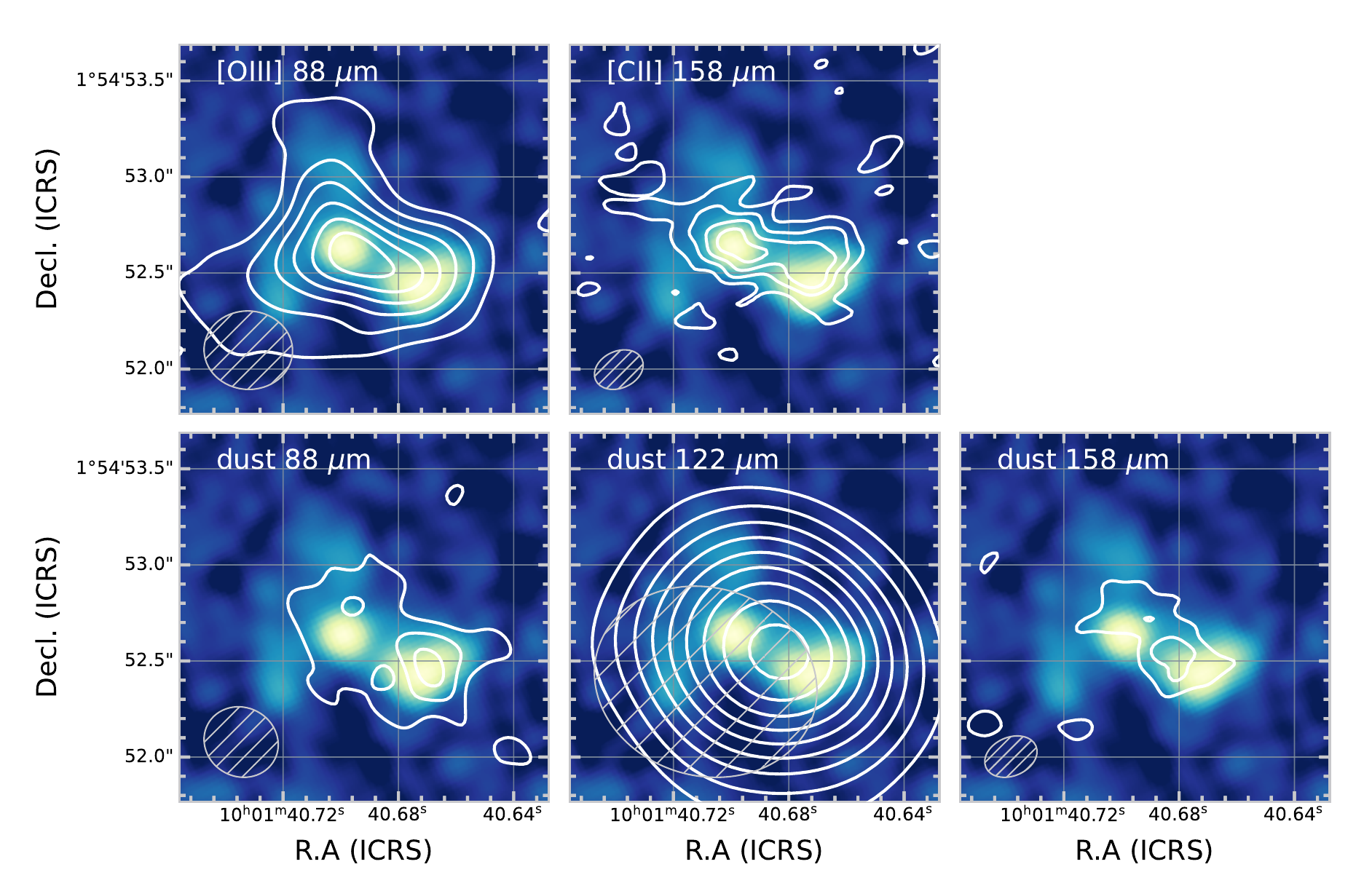}
    \caption{
      ALMA line and dust contours overlaid on the background NIRCam image.
      The NIRCam image is a stack of F115W to F356W bands to trace the continuum light, which is homogenized to the F444W PSF\@.
      The white contours are drawn with \( 2\sigma \) intervals from the \(2\sigma\) level, where \( \sigma = 0.04\) and \(0.02\) \Jybeamkms\ for the [\OIII] and [\CII] line maps and \(\sigma = 30\) and \(8\) \uJybeam\ for the \(88\) and \(158\) \um\ dust maps, respectively.
      For the \(122\) \um\ dust map, the white contours are drawn from the \(4\sigma\) level, where \( \sigma = 9\) \uJybeam.
      The gray hatched ellipses at the left bottom corners are the synthesized beam sizes of ALMA observations.
    }\label{fig:alma}
\end{figure*}

\begin{deluxetable*}{lcccccccccc}
    \tablecaption{NIRCam photometry and ALMA measurements of B14-65666 components.\label{tb:photometry}}
    \tabletypesize{\scriptsize}
    \tablewidth{0pt}
    \tablecolumns{11}
    \tablehead{
        \colhead{Aperture} & \colhead{F115W} & \colhead{F150W} & \colhead{F200W} & \colhead{F277W} & \colhead{F356W} & \colhead{F444W} & \colhead{Band8\(^{\dagger}\)} & \colhead{Band7\(^{\dagger\dagger}\)} & \colhead{Band6\(^{\dagger}\)} & \colhead{[OIII]88\(^{\dagger}\)}\\
        \colhead{} & \colhead{[$\mathrm{\mu Jy}$]} & \colhead{[$\mathrm{\mu Jy}$]} & \colhead{[$\mathrm{\mu Jy}$]} & \colhead{[$\mathrm{\mu Jy}$]} & \colhead{[$\mathrm{\mu Jy}$]} & \colhead{[$\mathrm{\mu Jy}$]} & \colhead{[$\mathrm{\mu Jy}$]} & \colhead{[$\mathrm{\mu Jy}$]} & \colhead{[$\mathrm{\mu Jy}$]} & \colhead{[$\mathrm{Jy~km~s^{-1}}$]}
    }
    \startdata
    Total & \(0.374\pm0.100\) & \(0.447\pm0.043\) & \(0.429\pm0.112\) & \(0.661\pm0.047\) & \(0.705\pm0.046\) & \(1.346\pm0.070\) & \(470\pm128\) & \(218\pm19\) & \(130\pm25\) & \(1.50\pm0.18\)\\
    E & \(0.159\pm0.060\) & \(0.162\pm0.023\) & \(0.261\pm0.051\) & \(0.343\pm0.027\) & \(0.355\pm0.024\) & \(0.773\pm0.040\) & \(208\pm83\) & - & \(41\pm23\) & \(0.92\pm0.14\)\\
    W & \(0.225\pm0.045\) & \(0.298\pm0.019\) & \(0.187\pm0.040\) & \(0.298\pm0.022\) & \(0.305\pm0.020\) & \(0.496\pm0.026\) & \(246\pm73\) & - & \(87\pm26\) & \(0.57\pm0.09\)\\
    E-core & \(0.107\pm0.022\) & \(0.128\pm0.008\) & \(0.163\pm0.018\) & \(0.180\pm0.011\) & \(0.167\pm0.010\) & \(0.440\pm0.022\) & - & - & - & -\\
    W-core & \(0.203\pm0.023\) & \(0.232\pm0.013\) & \(0.186\pm0.019\) & \(0.215\pm0.013\) & \(0.220\pm0.012\) & \(0.349\pm0.018\) & - & - & - & -\\
    N-tail & \(0.042\pm0.025\) & \(0.016\pm0.007\) & \(0.042\pm0.020\) & \(0.057\pm0.009\) & \(0.080\pm0.009\) & \(0.116\pm0.008\) & - & - & - & -\\
    S-tail & \(-0.005\pm0.038\) & \(0.014\pm0.011\) & \(0.050\pm0.033\) & \(0.091\pm0.014\) & \(0.091\pm0.011\) & \(0.161\pm0.011\) & - & - & - & -\\
    \enddata
    \tablerefs{\(^{\dagger}\)\citet{Hashimoto.T:2019a}, \(^{\dagger\dagger}\)\citet{Sugahara.Y:2021a}}
\end{deluxetable*}
 % tb:photometry
\vspace{-2\baselineskip}

Table~\ref{tb:photometry} lists measured photometry of the total and the galaxies E and W.
The table also lists photometry of the tails (N-tail and S-tail) and the cores of the galaxies E and W (E-core and W-core).
The applied apertures are shown in the left panel of Figure~\ref{fig:o3}.
Our measurements of the total photometry are consistent with the values on the COSMOS2020 catalog \citep[\textit{J}, \textit{H}, and \textit{Ks} bands in UltraVISTA and Ch1 and 2 in \textit{Spitzer};][]{Weaver.J:2022a}.
The F115W magnitudes of the galaxies E and W are measured to be \(25.9^{+0.5}_{-0.3}\) and \(25.5^{+0.2}_{-0.2}\) mag, respectively, corresponding to \(M_\text{UV} = -21.1^{+0.5}_{-0.3}\) and \(-21.5^{+0.2}_{-0.2}\) mag.
These magnitudes are similar to and brighter than the characteristic magnitude, \(M^{*}_\text{UV} \simeq -21.0\) mag, of the UV luminosity function at \(z\sim7\) \citep[e.g.,][]{Finkelstein.S:2015b, Bouwens.R:2021b}.
From UV to optical wavelengths, the galaxy E has a redder color than the galaxy W.
The tails largely contribute this red color of the galaxy E, as shown in Figure~\ref{fig:nircam}.
This color difference suggests that the galaxies E and W have different stellar and dust properties.

A characteristic color index is the F444W excess, which is twice as high as the F356W flux.
This excess corresponds to the [\OIII]\lamlam4960,5008 emission line strength.
While the \Hb\ line contributes to the F356W flux, both \Hb\ and [\OIII]4960,5008 lines increase the flux measured in the F444W filter.
The filter transmission of the \Hb\ line is similar in both of the bands, 37\% (F356W) and 50\% (F444W).
Therefore, by assuming a constant continuum flux density between F356W and F444W, we created a optical [\OIII] line map from the F356W\(-\)F444W color with little contamination of \Hb\ emission.
Other emission lines of \Hgamma, \Hdelta, and [\OIII] 4363 \AA\ may lead to underestimation of the [\OIII] flux by \(8\nn15\)\%, according to \textsc{Cloudy} calculations at \(0.1 < \Zgas/\Zsun < 0.4\) and \(-1.5 < \log{U} < -2.5\).
A possible red continuum slope would reduce \(\simeq\!10\)\% of the [\OIII] line flux and \(\simeq\!20\)\% of the [\OIII] equivalent width per the F356W\(-\)F444W color of \(0.1\)~mag.

The middle panel of Figure~\ref{fig:o3} shows the optical [\OIII] line map, which indicates strong [\OIII] emission at the cores.
The E-core and W-core exhibit integrated [\OIII] fluxes of \((4.3 \pm 0.4) \times10^{-17}\) and \((2.0 \pm 0.3) \times10^{-17}\) \ergscm, respectively.
Being regarded as a proxy of star formation, this strong [\OIII] emission indicates nuclear starbursts likely induced by the galaxy interaction, as seen in major mergers in the local Universe \citep[e.g.,][]{Ellison.S:2008a}.
The right panel of Figure~\ref{fig:o3} shows a map of the rest-frame [\OIII] equivalent width (\(\text{EW}_{0}\)).
The E-core and W-core exhibit \(\text{EW}_{0}([\OIII]) \sim 2050\pm290\) and \(740\pm170\) \AA, respectively, which are at the higher end of the distribution of [\OIII] emitters at \(z = 5\nn7\) \citep{Matthee.J:2023a, Boyett.K:2024a}.
In the tails, the [\OIII] fluxes are detected at the \(3\nn4\sigma\) significance levels, although this value may be overestimated because their red color may reflect their red optical continuum slopes, which is different from our assumption.
The obtained [\OIII] fluxes and equivalent widths are listed in Table~\ref{tb:properties}.

\subsection{NIRCam and ALMA imaging: Stellar versus dust distribution}\label{sec:result-alma}

The combination of the new NIRCam images with archival high angular-resolution (\(< 0\farcs4\)) ALMA data provides us with an unique opportunity to investigate the detailed stellar, ionized, and dust structure in an interacting system at \(z > 7\).
Figure~\ref{fig:alma} compares the NIRCam image, and ALMA line and dust maps.
The NIRCam image is a composite of all the bands, except for F444W, to exhibit the stellar continuum.
All the ALMA detections overlap the positions of the galaxies E and W, showing no spatial offsets between the UV-to-optical and FIR dust and line emission.
The [\CII] 158 \um\ and [\OIII] 88 \um\ emission lines in the galaxy E are \( 1\nn1.5\) times as bright as those in the galaxy W \citep{Hashimoto.T:2019a}.
This trend is qualitatively the same as the one seen in our [\OIII]\lamlam4960,5008 line map (Table~\ref{tb:properties}).
In contrast to the emission lines, the dust emission at \(88\) and \(158\) \um\ is brighter in the galaxy W than in the galaxy E \citep{Hashimoto.T:2019a}\footnote{The peak position of the \(158\) \um\ dust emission is shifted to the east of the galaxy W beyond the ideal astrometric uncertainty (Section \ref{sec:observational-data}). The origin of this shift is unclear, but it might be ALMA astrometric uncertainties poorer than the ideal one.}.
This dust spatial distribution might indicate that the galaxy W experiences dustier star-formation than the galaxy E.
As denoted in Section~\ref{sec:result-morphology}, however, the galaxy W presents a bluer UV color than the galaxy E, which is inconsistent with a previous finding that dust emission tend to be accompanied with red components in \(z > 6\) galaxies \citep{Bowler.R:2022a}.
We will come back to this point in Section~\ref{sec:irx-beta}.

The tails are detected only in the [\OIII] 88 \um\ line map, whereas they appear neither in the [\CII] 158 \um\ line map nor in the dust maps.
NIRCam detects the tails in the stacked continuum image, generated using the bands from F115W to F356W, and the [\OIII]\lamlam4960,5008 line map.
However, the tails are faint in the rest-frame UV filters (e.g., F150W), indicating red continuum slope from UV to optical.
The presence of continuum and line emission implies that the tails are actively forming stars, but the red continuum slope and the no emission of dust continua suggest a complex nature of the tails.

\begin{figure*}[t]
    \epsscale{1.2}
    \plotone{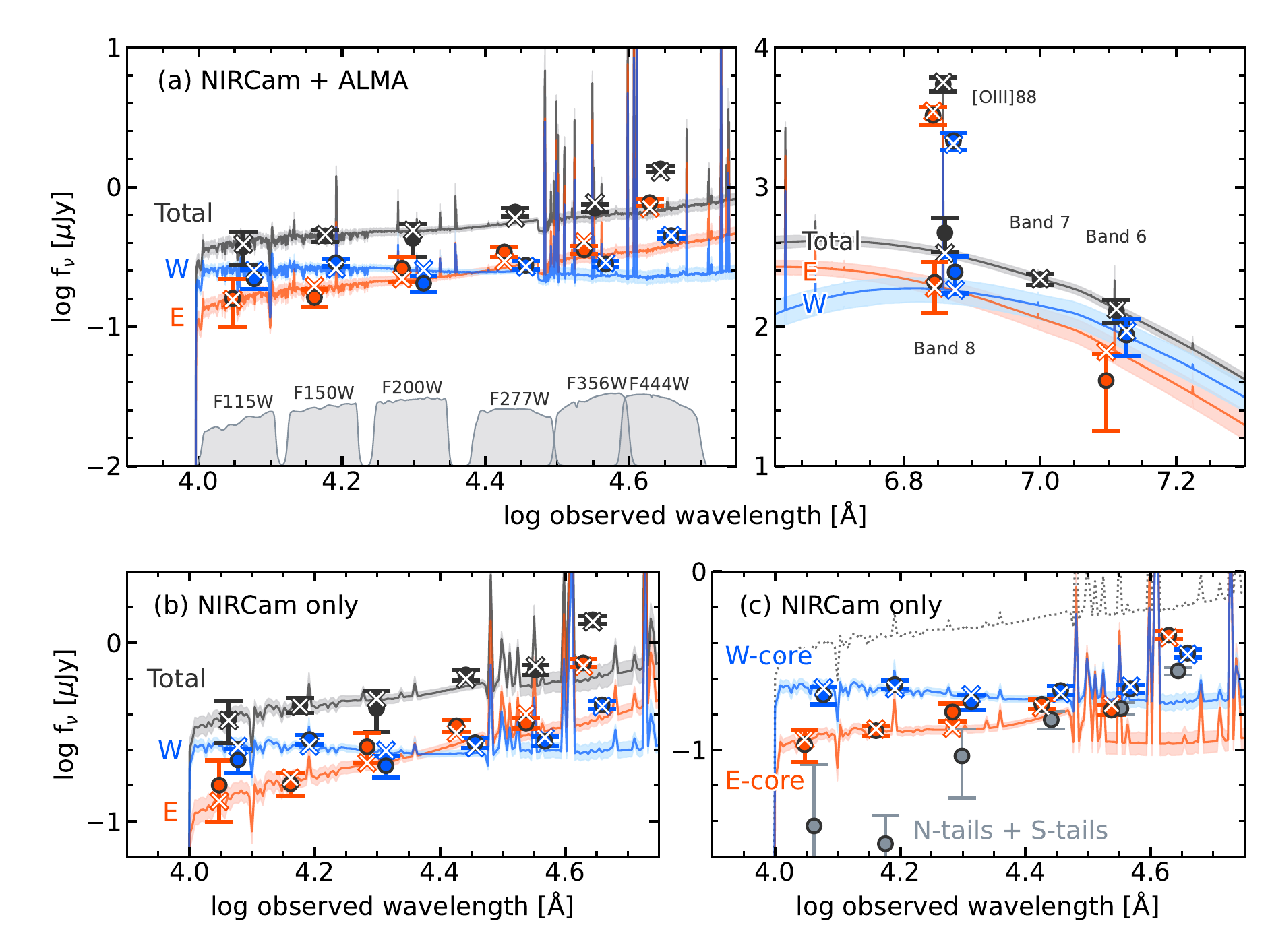}
    \caption{SEDs of B14-65666 and the results of the SED fitting.
      (a) The SED fitting results for the NIRCam and ALMA measurements.
      Each color shows the data of each component: total (black) and the galaxies E (red) and W (blue).
      The circles indicate the measured photometry and the error bars indicate the measurement errors including the absolute flux uncertainties.
      The data points are shifted from the centers of the filters for illustration purpose.
      The crosses show the obtained best-fit photometry of the SED fitting.
      The solid lines and shades show the best-fit curves and uncertainties taking \(16\) to \(84\) percentiles of the posterior distributions.
      The bottom gray shades of the left panel show the transmission curves of the NIRCam filters.
      (b) The SED fitting results only for the NIRCam photometry, without ALMA measurements.
      The best-fit parameters are comparable to the ones obtained for the NIRCam and ALMA measurements (Table~\ref{tb:sed}).
      (c) The SED of the E-core (red) and W-core (blue) and the obtained best-fit results for them.
      The gray data points show the SED of the tails, which is the sum of the N-tail and S-tail.
      The dotted line is shown for comparison, which is the same as the black line in the panel (b).
    }\label{fig:sed}
\end{figure*}

\begin{deluxetable*}{lccccccc}
    \tablecaption{Observed and Physical Properties of Components.\label{tb:properties}}
    \tabletypesize{\small}
    \tablewidth{0pt}
    \tablecolumns{8}
    \tablehead{
      \colhead{Components} & \colhead{Aperture diameter} & \colhead{Flux([\OIII])} & \colhead{EW\(_{0}([\OIII])\)} & \colhead{SFR\(_{0}\)} & \colhead{sSFR\(_{0}\)} & \colhead{SFR\(_\text{10Myr}\)} & \colhead{\(\beta_\text{UV}\)}\\
        \colhead{} & \colhead{[$\mathrm{arcsec^{2}}$]} & \colhead{[$\mathrm{10^{-17}\,erg\,s^{-1}\,cm^{-2}}$]} & \colhead{[$\mathrm{\mathring{A}}$]} & \colhead{[$\mathrm{M_{\odot}\,yr^{-1}}$]} & \colhead{[$\mathrm{Gyr^{-1}}$]} & \colhead{[$\mathrm{M_{\odot}\,yr^{-1}}$]} & \colhead{}
    }
    \startdata
    Total & \(1.8 \times1.8\) & \(10.1\pm1.3\) & \(1140\pm220\) & \(225^{+71}_{-56}\) & \(52^{+56}_{-28}\) & \(207^{+65}_{-51}\) & \(-1.57^{+0.11}_{-0.13}\)\\
    E & \(1.4 \times0.80\) & \(6.6\pm0.7\) & \(1480\pm260\) & \(178^{+52}_{-39}\) & \(53^{+26}_{-20}\) & \(178^{+50}_{-39}\) & \(-1.18^{+0.15}_{-0.16}\)\\
    W & \(1.1 \times0.66\) & \(2.6\pm0.4\) & \(750\pm170\) & \(37^{+12}_{-8}\) & \(23^{+15}_{-11}\) & \(37^{+12}_{-8}\) & \(-2.19^{+0.11}_{-0.09}\)\\
    E-core & \(0.44 \times0.44\) & \(4.3\pm0.4\) & \(2050\pm290\) & \(180^{+113}_{-58}\) & \(506^{+253}_{-182}\) & \(34^{+13}_{-9}\) & \(-1.78^{+0.11}_{-0.13}\)\\
    W-core & \(0.44\times0.44\) & \(2.0\pm0.3\) & \(740\pm170\) & \(29^{+13}_{-7}\) & \(27^{+28}_{-12}\) & \(29^{+14}_{-6}\) & \(-2.24^{+0.13}_{-0.10}\)\\
    N-tail & \(0.70\times0.40\) & \(0.6\pm0.2\) & \(580\pm260\) & - & - & - & -\\
    S-tail & \(1.0\times0.50\) & \(1.1\pm0.2\) & \(960\pm340\) & - & - & - & -\\
    \enddata
    \tablecomments{The flux([\OIII]) and EW\(_{0}([\OIII])\) are the line fluxes and rest-frame equivalent widths derived from the optical [\OIII]\lamlam 4960,5008 line map (Figure \ref{fig:o3}) within the component apertures. SFR\(_{0}\) is the instant star-formation rate (\(\SFR_0 = M_{*}/t_\text{SF}\)) and SFR\(_\text{10Myr}\) is the star-formation rate averaged in 10 Myr, that is, \(\SFR_\text{10Myr} = M_* (< 10\ \text{Myr})/10\ \text{Myr} \). The specific SFR\(_{0}\) is defined as \(1/t_\text{SF}\). \(\beta_\text{UV}\) is the UV spectral slope.}
\end{deluxetable*}
 % tb:tails
\vspace{-2\baselineskip}

\subsection{SED fitting of total and resolved components}\label{sec:result-sed-fitting}

We performed SED fitting combining the NIRCam and ALMA datasets for the entire system and the resolved galaxy components of B14-65666.
The apertures for each component is illustrated in the left panel of Figure~\ref{fig:o3} and the NIRCam and ALMA fluxes used in the SED fitting are summarized in Table~\ref{tb:photometry}.
The best-fit SEDs and measured physical parameters are presented in Figure~\ref{fig:sed} and Table~\ref{tb:sed}, respectively.
The best-fit parameters are given as \(2\sigma\) upper/lower limits when the peak of the posterior distributions is near the edge of the parameter boundaries (see Appendix \ref{sec:A1}).

\begin{deluxetable*}{lcccccccc}[h]
    \tablecaption{Results of the SED fitting.\label{tb:sed}}
    \tabletypesize{\small}
    \tablewidth{0pt}
    \tablecolumns{9}
    \tablehead{
        \colhead{Components} & \colhead{\(t_\mathrm{SF}\)} & \colhead{\(\log{M_*}\)} & \colhead{\(A_\mathrm{V}\)} & \colhead{\(\log U\)} & \colhead{\(Z_{*}\)} & \colhead{\(\log{M_\mathrm{old}}\)} & \colhead{\(\mathcal{U_\mathrm{min}}\)} & \colhead{\(\gamma\)}\\
        \colhead{} & \colhead{[$\mathrm{Myr}$]} & \colhead{[$\mathrm{\log M_{\odot}}$]} & \colhead{[$\mathrm{mag}$$\mathrm{\left( \mathrm{AB} \right)}$]} & \colhead{} & \colhead{[$\mathrm{Z_{\odot}}$]} & \colhead{[$\mathrm{\log M_{\odot}}$]} & \colhead{} & \colhead{}
    }
    \startdata
    \sidehead{NIRCam + ALMA data\tablenotemark{\(\dagger\)}}\
    Total & \(35^{+20}_{-12}\) & \(9.78^{+0.16}_{-0.18}\) & \(0.78^{+0.07}_{-0.06}\) & \(-2.09^{+0.58}_{-0.34}\) & \(0.21^{+0.05}_{-0.04}\) & \(<9.59\) & \(>10.45\) & \(0.88^{+0.07}_{-0.07}\)\\
    E & \(<93\) & \(9.55^{+0.19}_{-0.20}\) & \(0.98^{+0.08}_{-0.07}\) & \(-2.06^{+0.61}_{-0.34}\) & \(0.19^{+0.04}_{-0.04}\) & \(<9.55\) & \(>16.33\) & \(>0.83\)\\
    W & \(22^{+13}_{-8}\) & \(9.09^{+0.16}_{-0.18}\) & \(0.46^{+0.08}_{-0.06}\) & \(-1.95^{+0.54}_{-0.39}\) & \(0.32^{+0.06}_{-0.05}\) & \(<8.95\) & \(>7.61\) & \(<0.80\)\\
    \sidehead{Only NIRCam data\tablenotemark{\(\natural\)}}
    Total & \(18^{+21}_{-9}\) & \(9.61^{+0.25}_{-0.27}\) & \(0.86^{+0.11}_{-0.11}\) & \(-2.14^{+0.47}_{-0.34}\) & \(0.26^{+0.09}_{-0.08}\) & \(<9.68\) & - & -\\
    E & \(<53\) & \(9.51^{+0.17}_{-0.14}\) & \(1.21^{+0.13}_{-0.14}\) & \(-2.19^{+0.42}_{-0.30}\) & \(0.28^{+0.09}_{-0.08}\) & \(<9.28\) & - & -\\
    W & \(42^{+39}_{-17}\) & \(9.22^{+0.17}_{-0.17}\) & \(0.34^{+0.12}_{-0.09}\) & \(-2.08^{+0.54}_{-0.37}\) & \(0.24^{+0.09}_{-0.08}\) & \(<8.95\) & - & -\\
    E-core & \(<5\) & \(8.53^{+0.15}_{-0.13}\) & \(0.72^{+0.15}_{-0.14}\) & \(-1.82^{+0.49}_{-0.29}\) & \(0.33^{+0.11}_{-0.10}\) & \(<8.87\) & - & -\\
    W-core & \(35^{+29}_{-18}\) & \(9.02^{+0.19}_{-0.20}\) & \(0.30^{+0.14}_{-0.10}\) & \(-1.99^{+0.67}_{-0.38}\) & \(0.26^{+0.10}_{-0.10}\) & \(<8.89\) & - & -\\
    \enddata
    \tablenotetext{\dagger}{The results for the NIRCam and ALMA photometry.}
    \tablenotetext{\natural}{The results only for the NIRCam photometry. The dust emission parameters \(\mathcal{U_\text{min}}\) and \(\gamma\) are not used in the fits.}
    \tablecomments{The fitting parameters are described in Section \ref{sec:sed-fitting}. The best-fit values and \(1\sigma\) uncertainties are computed from a 50 percentile and 16 and 84 percentiles of the posterior distributions, respectively. The upper and lower limits are the \(2\sigma\) limits, that is, 97.5 and 2.5 percentiles of the posterior distributions, respectively.}
\end{deluxetable*}
 % tb:sed
\vspace{-2\baselineskip}

\subsubsection{Entire Merging System}\label{sec:whole-system-main}

The SED fit to the integrated flux supports that the B14-65666 system is a young starburst, regardless of whether including the ALMA data.
In the case without the ALMA data, the star-formation age and the stellar mass are \(t_\text{SF} = 20^{+20}_{-10}\) Myr and \(\log M_{*}/M_{\sun} = 9.6^{+0.3}_{-0.3}\), and then, instant star-formation rate is \(\SFR_0 = M_{*}/t_\text{SF} = 230^{+70}_{-60}\) \Moyr.
The ionization parameter is \(\log U = -2.1^{+0.5}_{-0.3}\).
This young starburst with the high ionization parameter gives rise to strong [\OIII] emission lines that can explain the F444W excess.
The relatively red stellar continuum even with the young starburst is attributed to the dust attenuation of \(A_{V} = 0.86^{+0.1}_{-0.1}\) mag.
The old stellar component is constrained to be \(\log M_{*}/M_{\sun} < 9.7\) at the \(2\sigma\) upper limit, which contributes the estimated continuum at F444W by up to \(10\)\%.
This upper limit is comparable to the stellar mass formed by the present star-formation.
Thus, the starburst induced by the major merger formed a stellar mass comparable to or higher than the old stellar mass built before the on-going merger.

The SED fitting including the dust continua shows higher \Umin\ and \(\gamma\) values than those of nearby galaxies \citep[\(\Umin \lesssim 10\) and \(\gamma \lesssim 0.1\) in][]{Draine.B:2007a}.
Such high \Umin\ and \(\gamma\) were also obtained for UV-selected galaxies at \(z > 4.5\) \citep{Burgarella.D:2022b}, indicating the necessity of high dust temperature to explain the FIR dust emission in high-redshift galaxies.
The characteristic dust temperature\footnote{Within the given fitting parameter ranges of \Umin\ and \(\gamma\), the maximum characteristic temperature is \(52.4\) K.}, derived from Equation (33) in \citet{Draine.B:2007a} and Equation (13) in \citet{Draine.B:2014a}, is \(T_\text{d, char} \simeq 18 \langle \mathcal{U} \rangle ^{1/6}\,\text{K} = 48^{+2}_{-3}\) K, which is consistent with the dust temperature obtained from the modified blackbody fitting \citep{Sugahara.Y:2021a}.
Although including the ALMA data slightly decreases the uncertainties of the best-fit dust attenuation and stellar metallicity, the changes in the best-fit stellar properties are insignificant.

Previous analyses of the entire properties of B14-65666 yielded stellar masses that are less than one-third of the one presented in this work \citep{Bowler.R:2018a, Hashimoto.T:2019a}.
On the other hand, the instant \(\SFR_0\) that they obtained are consistent with our estimate \citep{Hashimoto.T:2019a}.
The stellar mass discrepancy is because of the differences on the IRAC and NIRCam photometry, where IRAC flux densities were \(1.5\nn2\) times as low as the NIRCam flux densities at \(3.6\) and \(4.4\) \um.
We note again that our new photometry agrees with that in the COSMOS2020 catalog, which is also higher than the previous photometry at \(3.6\) \um\ and \(4.5\) \um.

\subsubsection{Resolved components}\label{sec:resolved-components}

The high angular resolution of the NIRCam imaging enables us to perform the SED fitting to each of the galaxy components.
Figure~\ref{fig:sed} makes it clear that the galaxies E and W have different colors in the rest-frame UV-to-optical wavelengths.
In the cases without the ALMA data, they are inferred as the young starbursts with ages of \( < 53\) Myr and \(40^{+40}_{-20}\) Myr, respectively, similar to the total.
The \(\SFR_{10 \text{Myr}}\ ( = M_{*}( < 10\, \text{Myr})/10\, \text{Myr})\) of the galaxy E (\(180^{+50}_{-40}\) \Moyr) is \(\simeq4\) times as high as that of the galaxy W (\(40^{+10}_{-10}\) \Moyr).
Moreover, the galaxy E is expected to have \(\gtrsim0.5\) mag stronger dust attenuation than the galaxy W.
These results reflect a dusty, bursty phase of the galaxy E, which is also supported by the strong F444W excess, i.e., the highest [\OIII] line strength at the core of the galaxy E (Section~\ref{sec:result-morphology}).
In the SED fitting including the ALMA data, the [\OIII] 88 \um\ emission line fluxes help more strictly constrain the metallicity of the galaxies; the galaxy E exhibits a lower stellar metallicity than the galaxy W while they have a similar metallicity in the SED fitting without the ALMA data.
These metallicity constraints originate from better constraints on the gas-phase metallicity by the optical-to-FIR [\OIII] line ratios.
Both of the galaxies show small Balmer jumps (or inverse Balmer break), a flux excess at wavelengths shorter than \(4000\) \AA\ caused by the nebular continuum, which is especially strong for the galaxy E.
The contribution of this nebular continuum and [\OII]\lamlam3727,3730 emission lines help to boost the F277W flux at a level similar to that of the F356W filter.
The sum of the stellar masses of the galaxies is \(\log M_{*} / M_{\sun} \simeq 9.68\), which is comparable to the stellar mass derived from the total photometry.
We note that the old stellar component (\(M_\text{old}\)), which is frequently missed in modeling only with young starburst populations \citep{Roberts-Borsani.G:2020b, Gimenez-Arteaga.C:2023a}, may contribute up to a factor of two to the total stellar mass as discussed in the previous section.
Finally, the stellar-mass ratio of the galaxies E and W is estimated to be 3:1 to 2:1.
This mass ratio is comparable to the dynamical mass ratio of the galaxy E (\(5.7\pm1.6\times10^{10}\) \Mo) to the galaxy W (\(3.1\pm1.1\times10^{10}\) \Mo) that are inferred from the [\OIII] 88 \um\ line widths on the assumption of the virial theorem \citep{Hashimoto.T:2019a}.
Thus, this study has revealed that B14-65666 is surely a major merger in terms of the stellar mass ratio of the galaxies, while previous studies identified it from the luminosity ratio.

In the FIR SED, it is notable that the galaxies E and W have different peaks of the dust continua.
The NIRCam and ALMA observations have shown that the galaxy E has redder UV spectral slope while the galaxy W exhibits higher dust-continuum flux densities.
The SED fitting explains these observations with a difference of the dust temperatures.
In particular, the galaxy E requires so high \Umin\ and \(\gamma\) that their probability distribution functions reach the maximum values of their permitted parameter ranges.
The characteristic dust temperatures are \(T_\text{d, char} = 50^{+1}_{-2}\) and \(40^{+4}_{-3}\) K for the galaxies E and W, respectively.
We will further discuss relations between the UV and FIR light in Section~\ref{sec:irx-beta}.

For finer resolved analyses, we divided the system into the cores of the galaxies and the tails in the galaxy E.
We did not include the ALMA data in these SED fittings, which have relatively lower angular resolution.
The panel (c) of Figure~\ref{fig:sed} depicts the observed SED and best-fit spectra for the E-core and W-core.
The W-core has a similar SED shape to the galaxy W and is \(10\nn30\)\% fainter than it; this similarity means that the W-core is dominant in the SED of the galaxy W\@.
On the other hand, the E-core is twice as faint as the galaxy E at the long wavelength channels.
The obtained best-fit age is very young, \( < 5\) Myr, at the \(2\sigma\) upper limit and the stellar mass is \(\simeq1.0\) dex lower than that of the galaxy E.
As shown in the panels (b) and (c), there are a \( > 0.4\) dex difference in the best-fit stellar continua at wavelengths longer than the Balmer jump.
These differences indicate that the red tails strongly affect the SED of the galaxy E at the rest-frame optical wavelengths.
One possible scenario to interpret this result might be that the galaxy E is composed of the dusty compact starbursts at the nucleus and the older stars in the tails which dominate the stellar mass.
However, the non-detections in the F115W, F150W, and F200W bands of the tails make it challenging to understand the origin of their red color by solving the degeneracy between the old stellar populations, dust attenuation, and strong emission lines (Figure \ref{fig:sed}).
Detailed properties of the tails will be estimated by including emission lines taken in approved JWST NIRSpec and MIRI observations.

\section{Discussion}\label{sec:discussion}

\begin{figure}[t]
    \epsscale{1.2}
    \plotone{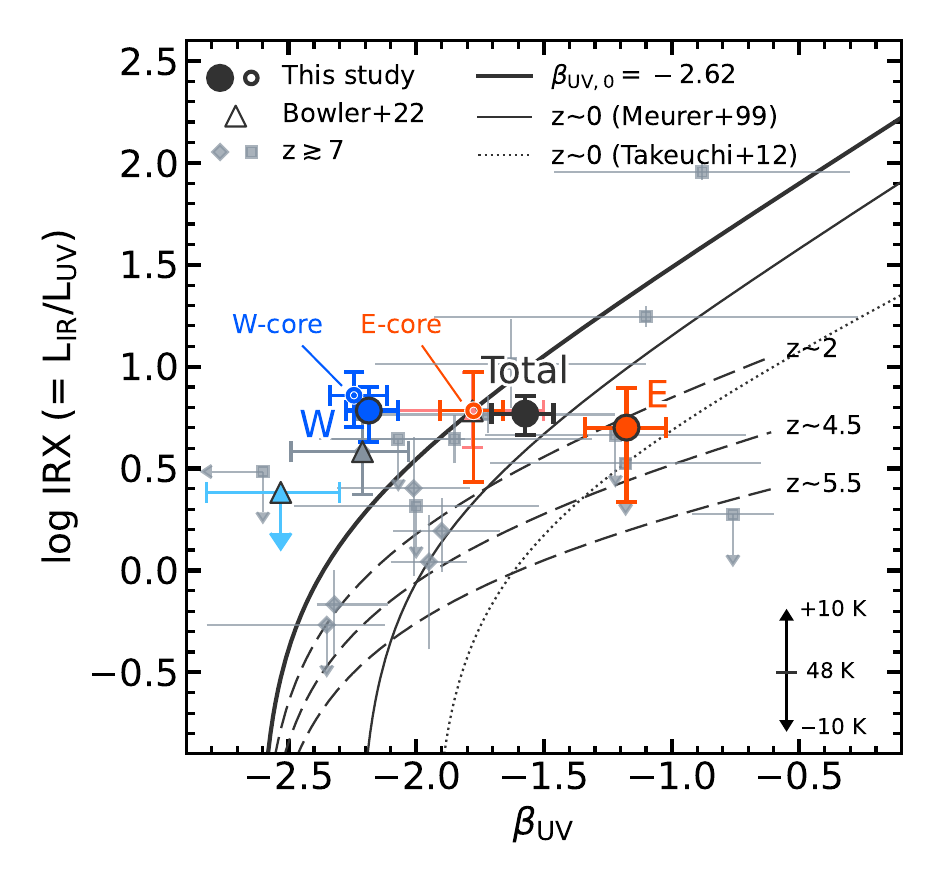}
    \caption{IRX--\betaUV\ diagram.
      The black, red, and blue circles show the measurements for the total, galaxy E, and galaxy W of B14-65666.
      The E-core and W-core are depicted with the small circles with the same colors of the galaxies E and W, respectively.
      The light-color triangles show the measurements for B14-65666  obtained by \citet{Bowler.R:2022a}.
      We note that the galaxy-E symbol of Bowler et al.\ is completely overlapped with our E-core symbol.
      The small gray symbols show \(z\sim7\) galaxies taken from \citet[][diamond]{Bowler.R:2022a} and \citet[][square]{Hashimoto.T:2019a}.
      All the IRX values of the data points from the literature are corrected for \(T_\text{d} = 48\) K and \(\beta_\text{d} = 2.0\).
      The lines show the parameterized IRX--\betaUV\ relations:
      the black solid line is the \citet{Meurer.G:1999a} relation modified with \(\betaUVO = -2.62\) \citep{Reddy.N:2018a},
      the thin solid line is the original Meurer relation (\(\betaUVO = -2.23\)) for UV-selected local starbursts,
      the dotted line is an aperture-corrected Meurer relation \citep{Takeuchi.T:2012a},
      and the dashed lines are relations obtained for UV-selected galaxies at \(z\sim2\) \citep{Reddy.N:2018a} and \(z\sim4.5\) and \(5.5\) \citep{Fudamoto.Y:2020a}.
      The bottom right arrows show how the IRX values change if the assumed dust temperature increases or decreases from \(48\) K.
    }\label{fig:irxb}
\end{figure}

\subsection{IRX--\(\beta_\text{UV}\) relation}\label{sec:irx-beta}

The NIRCam and ALMA data presented so far have demonstrated the complex stellar and ISM structure of the system B14-65666.
% The NIRCam and ALMA images have indicated complex morphology of B14-65666 in the rest-frame UV, optical, and FIR views.
While the galaxy W is bluer than the galaxy E in the rest-frame UV (Figure~\ref{fig:sed}), it exhibits brighter dust emission (Figure~\ref{fig:alma}).
This originates from different properties of star formation and dust attenuation between the galaxies E and W.
The SED fitting in Section \ref{sec:result-sed-fitting} has suggested the different dust temperature as the origin, but there are other physical mechanisms including assumptions of the dust attenuation curve.

We investigate the relation between the infrared excess (\(\text{IRX} = L_\text{IR}/L_\text{UV}\)) and the UV spectral slope \betaUV\ to study whether the discrepancy between the UV colors and dust emission is due to the dust attenuation curve.
We computed the monochromatic rest-frame UV luminosity \(L_\text{UV}\) and \betaUV\ from the best-fit SED models derived without the ALMA data.
The UV luminosity was computed from the UV continuum as \(L_\text{UV} = \lambda L_{\lambda}\) at \(\lambda = 1600\) \AA\@.
The UV slope was obtained by linear fitting to the UV continuum within the wavelength windows proposed by \citet{Calzetti.D:1994a}.
The measurement uncertainties were propagated from the fitting uncertainties; we measured values from the 500 SED models derived from the parameter posterior distributions and estimated uncertainties from the measurement distribution.
The IR luminosity \(L_\text{IR}\) was taken from \citet{Hashimoto.T:2019a}, who estimated it by the modified blackbody fitting under the assumption of the dust temperature of \(T_\text{d} = 48\) K and the dust emissivity of \(\beta_\text{d} = 2.0\).
This assumed dust temperature agrees with the characteristic dust temperature \(T_\text{d, char}\) of the entire system that has been derived in the SED fitting.

The obtained IRX values of the entire system (total) and both of the galaxies are similar, \(\log \text{IRX} \simeq 0.7\).
However, the \betaUV\ values range from \(\betaUV = -2.2\) (galaxy W) to \(-1.2\) (galaxy E).
The UV slope \betaUV\ depends on stellar and dust properties including age, dust composition, and the stellar-and-dust geometry.
Therefore, the variation in \betaUV\ at a given IRX supports a spatial variation of stellar and dust properties between the components.

Previously \citet{Bowler.R:2022a} measured the IRX and \betaUV\ values for B14-65666.
In Figure~\ref{fig:irxb}, we plotted their data points after correcting the IRX values for \(T_\text{d} = 48\) K and \(\beta_\text{d} = 2.0\).
Compared with their values, our measurements are systematically redder in \betaUV\@.
Although the origin of the systematics is unclear, the \betaUV\ difference would be caused by different spatial resolution.
\citet{Bowler.R:2022a} used UltraVISTA images with the spatial resolutions of  \(\sim 0\farcs7\), much coarser than those of our NIRCam images, and they resolved objects with \texttt{TPHOT} \citep{Merlin.E:2015a} based on an HST image.
Thus, unexpected systematics may be included in their rest-frame UV photometry.
This comparison underlines the power of high resolution for spatially resolved studies.

The IRX--\betaUV\ relations are connected with attenuation curves under the energy balance between absorption and re-radiation by dust \citep{Meurer.G:1999a}.
Canonical relations are parameterized with attenuation curves, for example, for local starbursts \citep{Meurer.G:1999a, Overzier.R:2011a, Takeuchi.T:2012a} and for the Small Magellanic Cloud \citep[SMC;][]{Prevot.M:1984a, Gordon.K:2003a}.
For high-redshift galaxies, the intrinsic UV slope, \betaUVO, is bluer than that for the local starbursts \citep{Reddy.N:2018a} due to the young ages, low stellar metallicities, and binary interactions \citep{Steidel:2014,Steidel:2016}, although the nebular continuum emission could make it redder.
We checked \betaUVO\ of our best-fit intrinsic SED curves before dust reddening and found that it ranges from \(\betaUVO = -2.6\) to \(-2.55\), which is bluer than \(-2.23\) used in \citet{Meurer.G:1999a}.
For simplicity we use \(\betaUVO = -2.62\) \citep{Reddy.N:2018a} as a fiducial value at high redshift.

Figure~\ref{fig:irxb} compares the measurements for B14-65666 and various IRX--\betaUV\ relations.
The total is in a good agreement with the widely-used canonical relation, the \citet{Meurer.G:1999a} relation, modified to \(\betaUVO = -2.62\).
From this result, (1) the agreement with the Meurer relation supports our assumption of using the \citet{Calzetti:2000} attenuation curve in the SED fitting and (2) \(\betaUVO = -2.62\) supports lower stellar metallicity and younger ages of B14-65666 than those of local starbursts \citep{Meurer.G:1999a, Overzier.R:2011a, Takeuchi.T:2012a}. 
The galaxies E and W are, however, not consistent with the \betaUVO-modified Meurer relation.
The galaxies E and W exhibit lower and higher IRX than the relation at their \betaUV, respectively.
Deviations from the canonical IRX-\betaUV\ relations arise from several physical processes \citep[e.g.,][]{Popping.G:2017a}.
For the galaxy E, the low IRX value would be explained by hypotheses that the galaxy E has higher dust temperatures of \(T_\text{d} \ge 63\nn68\) K in the modified blackbody, as suggested by the SED fitting (Section~\ref{sec:resolved-components}); or that the galaxy E exhibits a steeper SMC-like dust attenuation curve than the \citet{Calzetti:2000} one.
Alternatively, the measurement of \betaUV\ based on the SED fitting may be affected by the red tails.
If we assume that the measured dust emission only arises from the E-core, the E-core is located on the \betaUVO-modified Meurer relation on the IRX--\betaUV\ diagram.
For the galaxy W, hypotheses to explain the IRX values are a low dust temperature of \(T_\text{d} \le 27\nn33\) K, which qualitatively agrees with the SED-fitting result; or the patchy stellar-and-dust geometry unresolved even with NIRCam and ALMA (\( \lesssim 0.2\nn1.5 \) kpc), which produces flatter dust-attenuation curves resulting in high IRX at blue \betaUV\@.
The patchy stellar-and-dust geometry has been already found in a Lyman-break galaxy at \(z = 8.31\) \citep{Tamura.Y:2023b} and supported by the FirstLight simulations, where complex stellar-and-dust geometry produces high IRX \citep{Mushtaq.M:2023a}.
In any scenarios, or in a combination of several scenarios, our rest-frame UV and FIR observations have highlighted the different dust properties between the galaxies E and W.

B14-65666 and most other \(z\gtrsim7\) galaxies \citep{Hashimoto.T:2019a, Bowler.R:2022a} are distributed between the Meurer relations with \(\betaUVO = -2.23\) (black thin line in Figure~\ref{fig:irxb}) and \(-2.62\) (black thick line).
From this distribution, \citet{Hashimoto.T:2019a} and \citet{Bowler.R:2022a} concluded that dust-detected galaxies at \(z\gtrsim7\) follow the \citet{Calzetti:2000} attenuation curve.
The apparent IRX dispersion would be explained by a combination of a \betaUVO\ variation (i.e., a variation of stellar age and metallicity) and patchy stellar-and-dust geometry \citep{Reddy.N:2018a}.
We note that their IRX is corrected for \(T_\text{d}\) and \(\beta_\text{d}\) used in this study and that their \betaUV\ was derived from photometry, which is different from our method.
These results are consistent with stacking analyses of Lyman-break galaxies at \(z\sim3\) \citep[e.g.,][]{Alvarez-Marquez.J:2016a, Alvarez-Marquez.J:2019b, Bowler.R:2024a}, but seem to be inconsistent with steeper attenuation curves favored in stacking analyses of UV-selected galaxies at \(z\sim2\) \citep{Reddy.N:2018a} and \(z\sim4.5\) and \(5.5\) \citep{Fudamoto.Y:2020a}.
However, some literature galaxies in Figure~\ref{fig:irxb} have stringent IRX upper limits below the canonical relations at high \betaUV\ and the IRX estimates significantly depend on the assumptions of \(T_\text{d}\) and \(\beta_\text{d}\) and how to measure \betaUV\@.
The ``standard'' IRX--\betaUV\ relation at high redshift is still under debate.

\begin{figure}[t]
    \epsscale{1.2}
    \plotone{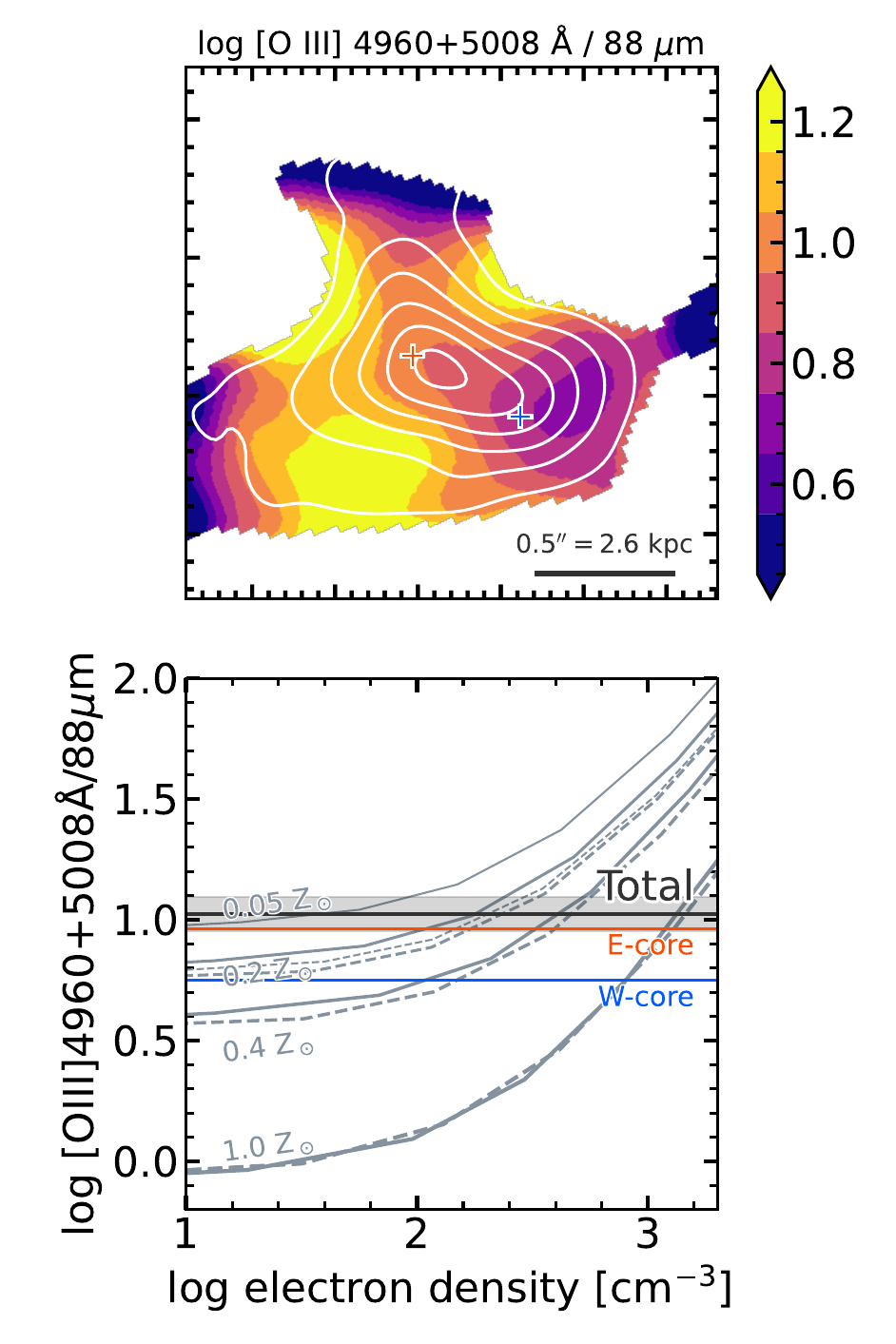}
    \caption{Top: Map of the optical-to-FIR [\OIII] emission line ratio, color coded by \(\log [\OIII]\ 4960 + 5008\ \mbox{\AA}/[\OIII]\ 88\ \mbox{\um}\).
      The white contours show the [\OIII] 88 \um\ fluxes, which is the same as in Figure~\ref{fig:alma}.
      The red and blue crosses depict the peak positions of the E-core and W-core, respectively.
      Bottom: the optical-to-FIR [\OIII] emission line ratio as a function of the electron density.
      The horizontal lines show the line ratios of the entire system (black) and the value at the peak positions of the E-core (red) and W-core (blue).
      The gray curves illustrate theoretical predictions computed from \textsc{Cloudy}.
      The line thicknesses depend on the gas-phase metallicity and the solid and dashed lines are at \(\log U = -1.5\) and \(-2.5\), respectively.
      At a typical electron density (\(\sim 200\nn400\) \cmmm), the gas-phase metallicity of the B14-65666 is \(\simeq 0.2\nn0.4\) \Zsun.
    }\label{fig:o3ratio}
\end{figure}

\subsection{[\OIII] Line Ratio: Insights into Gas-phase Metallicity}\label{sec:estimated-line-maps}

NIRCam images have identified a flux excess in the F444W band interpreted as a strong [\OIII]\lamlam4960,5008 emission.
The ALMA Band 8 has detected the [\OIII] 88 \um\ emission.
Due to the large difference in their excitation energy, the ratio of the optical to FIR [\OIII] emission lines is sensitive to the physical conditions (electron density and temperature) of the ionized emitting gas \citep{Osterbrock.D:2006a}.
This line ratio also depends on the gas-phase metallicity under the assumptions of photoionization models \citep[e.g.,][]{Nakazato.Y:2023a}.
Thus, a combination of the NIRCam images and ALMA moment-0 map provides us with insights into spatial distributions of gas-phase metallicity.

We created a line-ratio map by taking a ratio of the optical [\OIII]\lamlam4960,5008 emission-line map to the FIR [\OIII] 88 \um\ image.
The optical [\OIII] line map was created as in Section~\ref{sec:result-morphology}, but here the F356W and F444W images were corrected for the dust attenuation before creating the line map.
We corrected flux densities within the apertures E (\(A_\text{V} = 1.2\) mag), W (\(0.34\) mag), and E-core (\(0.72\) mag) so that the flux density recovered in each aperture agreed with the best-fit \(A_\text{V}\) value obtained in the SED fitting.
The created attenuation-corrected optical line map was convolved to match the angular resolution of the ALMA synthesized beam (\(0\farcs38\)).
The map area is limited to the region where both lines are detected at more than 1\(\sigma\) level and it corresponds to \(3\nn4\) beam solid angle.

The top panel of Figure~\ref{fig:o3ratio} illustrates the optical-to-FIR [\OIII] line-ratio map.
The line ratio at the peak of the [\OIII] 88 \um\ surface brightness is \(\simeq7.8\), and \(10.6\pm1.9\) for the integrated flux, where the error does not include the uncertainty of \(A_\text{V}\).
The figure shows the spatial variation of the line ratio over the galaxies, which suggests a variation of the electron temperature and/or electron density in the galaxies E and W.
At the peak positions of the E-core and W-core, the [\OIII] line ratios are \(9.2\) and \(5.6\), respectively.
The highest line ratios are seen in the S-tail; however, these high values are based on the assumed strong dust correction and include large uncertainties.

To be more quantitative, we used \textsc{Cloudy} to explore nebular parameters that reproduce the obtained [\OIII] line ratios.
In our photoionization models, the [\OIII] line ratio depends on nebular parameters: the gas-phase metallicity \Zgas, ionization parameter \(U\), and hydrogen density \(n_\text{H}\).
As the F356W band includes weak emission lines, which are also subtracted from F444W flux, we corrected the model [\OIII]\lamlam4960,5008 line fluxes by subtracting expected \Hgamma\ and \Hdelta\ line fluxes.
The bottom panel of Figure~\ref{fig:o3ratio} shows optical-to-FIR [\OIII] line ratios as a function of the nebular parameters.
The illustrated model lines depend on \Zgas\ and \(U\); low gas-phase metallicity and high ionization parameter increase the electron temperature of the ionized gas, leading to strong [\OIII]\lamlam4960,5008 lines with respect to [\OIII] 88 \um\ line.
Instead of \(n_\text{H}\) at the illuminated surface, we used the averaged electron density \(n_\text{e}\) for the \(x\)-axis that was computed from the modeled [\OIII] 52 and 88 \um\ line ratios at the electron temperature of \(1.2\times10^{4}\) K using PyNeb \citep{Luridiana.V:2015a}.

The estimated gas-phase metallicity is \(\Zgas\sim 0.2\nn0.4\) \Zsun\ for the total, at an assumed electron density of \(n_\text{e} \sim 200\nn400 \) \cmmm, which is a typical estimate from optical-to-FIR [\OIII] line ratios at high redshift \citep{Stiavelli.M:2023a, Fujimoto.S:2024b}.
Using the line ratios at the peak positions of the E-core and W-core, the E-core exhibits a lower gas-phase metallicity of \(Z\sim 0.2\nn0.4\) \Zsun\ than the W-core (\( > 0.4\) \Zsun).
These results are consistent with conclusions of the previous studies using the [\OIII] 88 \um\ emission line that B14-65666 has sub-solar metallicity but not extremely low at \( < 0.1\) \Zsun\ \citep{Hashimoto.T:2019a, Jones.T:2020a, Yang.S:2020a}.
The estimated values here are slightly higher than the metallicities derived from the SED fitting to the NIRCam and ALMA data, reflecting the assumed hydrogen (i.e., electron) density of \(n_\text{H} = 100\) \cmmm\ in \texttt{Bagpipes}.
We note that adopting higher electron densities of \(n_\text{e} \sim 600\) \cmmm, which is derived from optical emission lines \citep[e.g.,][]{Abdurrouf:2024a}, results in higher \Zgas.
In terms of the electron density, the optical-to-FIR [\OIII] line ratios disagree with the presence of a high electron density gas (\(n_\text{e} > 10^4\) \cmmm), which have been traced by high-ionization UV emission lines in some high-redshift galaxies \citep[e.g.,][]{Senchyna.P:2023a, Marques-Chaves.R:2024a}, under the assumption that galaxies in the EoR would have sub-solar metallicity.

\subsection{Overall scenario for B14-65666 system}\label{sec:picture-b14-65666}

The NIRCam and ALMA observations have shed light on the morphology and physical properties of B14-65666 in the EoR.
We analyzed the rest-frame optical and FIR data to show the complex morphology of the merging galaxies: the galaxy E is composed of the compact core with faint dust emission and the diffuse red tails; the galaxy W presents an elongated and clumpy morphology with bright dust emission.
These clumpy, elongated structures and diffuse emission like tidal tails are reproduced in the simulations of high redshift galaxies \citep[\(z = 6\nn9\),][]{Nakazato.Y:2024a}.
Furthermore, the galaxy W may exhibit patchy stellar-and-dust geometry that is not resolved in our observations.
These results indicate that B14-65666 is experiencing strong tidal forces accompanied with a merger, disturbing the entire morphology of the system and each of the galaxies involved in the process.

The projected separation of the galaxies E and W is \(d\simeq4\) kpc and their line-of-sight velocity difference is \(v\simeq200\) \kms\ \citep{Hashimoto.T:2019a}.
Although this is the first-order estimate, the timescale elapsed from the previous passage is estimated to be \(d/v\simeq20\) Myr, which is comparable to the star-formation age inferred from the SED fitting.
This merging timescale is feasible because it is shorter than the expected merger observability timescale of \(180\) Myr at \(z = 7.15\) \citep{Snyder.G:2017a}.
This consistency supports that the gravitational interactions have driven the star formation in the B14-65666 after the first (or second) passage.

Major mergers would promote gas accretion by removing the angular momentum from disk gas, triggering starbursts at the cores of the galaxies \citep[e.g.,][]{Mihos.J:1994a}.
B14-65666 shows signatures of nuclear dusty starbursts, including strong optical [\OIII] emission, strong dust attenuation, and high \SFR\@.
These nuclear dusty starbursts have indeed enhanced the star-forming activity of B14-65666.
The instant \(\SFR_0\) of \(230\) \Moyr\ is \(\simeq1\) dex higher than the star-formation main sequence at \(\log M_{*}/\Msun = 9.7\) and \(z\sim7\) \citep{Popesso.P:2023a}.
The merger-induced star-formation at high redshift is consistent with a finding in the FirstLight simulations that bursty star-formation is induced by gas-rich mergers at \(z > 5\) \citep{Ceverino.D:2018a}.
While many observations demonstrate that major mergers enhance \SFR\ in the local Universe \citep[e.g.,][]{Ellison.S:2008a}, the \SFR\ enhancement is reported to be inefficient at \(z > 1\) \citep[e.g.,][]{Silva.A:2018a, Shah.E:2022a}.
B14-65666 is a clear example experiencing enhanced star formation at higher redshift.

In particular, the E-core experiences a compact, centrally concentrated, strong starburst.
Given \(r_\text{e} < 85 \) pc of the E-core in the F115W image and \(\SFR_0\) values in Table~\ref{tb:properties}, the \(\SFR_0\) surface density of the E-core is \(\Sigma_{\SFR, 0}\gtrsim4\times10^3\) \Moyrkpct, which is comparable to IR \SFR\ surface densities of local ULIRGs or obscured AGN \citep{Pereira-Santaella.M:2021a}.
We expect that this compact starburst is caused by a significant gas accretion to the E-core.
The high \(\Sigma_{\SFR, 0}\) is also consistent with the expected high dust temperature of the E-core \citep[see][]{Liang.L:2019a}.
We note that star-formation timescale of the E-core (\( < 5\) Myr) is much shorter than those of ULIRGs, as \SFR\ of ULIRGs are derived from the FIR luminosity (i.e., \(\sim100\) Myr).
Although the molecular gas mass is highly uncertain for B14-65666 \citep[\(10^{8.7\nn11}\) \Msun,][]{Hashimoto.T:2023b}, the high \(\Sigma_{\SFR, 0}\) would lead to short gas depletion time of \(< 10\) Myr in the range of molecular gas mass surface densities of local starbursts \citep[\( < 10^4\) \Mopcpc;][]{Hodge.J:2015a, Pereira-Santaella.M:2021a}.
In addition, the high \(\Sigma_{\SFR, 0}\) and s\(\SFR_0\) of the E-core suggest that the radiation pressure is high enough to launch galactic outflows \citep[e.g.,][]{Pereira-Santaella.M:2021a, Ferrara.A:2023a}, which lead to negative feedback, therefore reducing or even quenching the star formation \citep{Hopkins:2012}.
There might also exist an AGN in the compact E-core, which would be consistent with its high surface brightness and high dust temperature.
In such a case, the E-core star-formation history will be affected by additional negative feedback by AGN.
Thus, the compact dusty starburst at the E-core may cease within several tens Myr.

The gas-phase metallicity for the entire system is estimated to be \(\simeq 0.2\) \Zsun\ from the SED fitting, and these estimates are supported by the optical-to-FIR [\OIII] emission line ratio.
B14-65666 is on the average mass-metallicity relation at \(z = 4\nn9\) \citep{Nakajima.K:2023a, Curti.M:2023a}, but given its high \SFR, it would not be on the fundamental metallicity plane \citep{Andrews.B:2013a}.
Our analyses also show that the galaxy E has lower metallicity than the galaxy W, which is consistent with a scenario where a strong accretion of less-enriched gas dilutes the pre-existing gas at the galaxy center \citep[e.g.,][]{Kewley.L:2006a}.
At \(\Zgas\simeq0.2\) \Zsun, an ionization parameter estimated by \citet{Sugahara.Y:2021a} using the [\OIII] 88 \um\ and [\CII] 158 \um\ lines is \(\log U \simeq - 2.0\), which is consistent with the SED fitting results.
In other words, high [\OIII]\(/\)[\CII] ratio of B14-65666 can be naturally explained by the high ionization parameter \citep[see][]{Hashimoto.T:2019a, Harikane.Y:2020b}.
The [\NII] 122 \um\ non-detection is also easily explained by the high ionization parameter where most of the nitrogen are in the doubly ionized phase \citep{Sugahara.Y:2021a}.

Finally, we mention some limitations of our observations.
Although we have revealed many properties of the cores of the galaxies, less properties are constrained for the tails.
This is mainly because it is difficult to solve the degeneracy in physics that can explain their red SED, dust non-detections, and detection of [\OIII] 88 \um\ emission, with different angular resolutions of the NIRCam and ALMA\@.
We expected that the extended diffuse emission originates from the tidal tails produced by the major merger, but we did not reject other possibilities like ionized gas outflows \citep[e.g.,][]{Yuma.S:2019a} and offset old stellar populations \citep[e.g.,][]{Colina.L:2023a}.
Approved MIRI and NIRSpec IFU spectroscopy and ALMA high angular-resolution observations will give a hint to properties of the tails.
To reach statistical conclusions, a large sample of major mergers at high redshift is necessary.
It is unclear whether B14-65666 is representative of major mergers at this redshift.
Nevertheless, our results suggest that the major merger, which is an abundant population at high redshift \citep{Rodriguez-Gomez.V:2015a, Duncan.K:2019a, Romano.M:2021a}, makes significant contributions to galaxy mass assembly in the early Universe.
B14-65666 is still an important target that provides us with detailed distributions of stellar, gaseous, and dust populations during major mergers.

\section{Conclusions}\label{sec:summary}

In this paper, we present JWST NIRCam observations of the UV-bright (\(M_\text{UV} = -22.5\) mag) Lyman-break galaxy system B14-65666 at \(z = 7.1520\), also known as Big Three Dragons.
The NIRCam filters are selected to sample the rest-frame UV wavelengths, the Balmer break clean of strong optical emission lines, and the rest-frame optical wavelengths (\(\sim5000\) \AA) including \Hb\ and [\OIII] emission lines.
Additionally, rich ancillary ALMA data trace the [\OIII] 88 \um\ and [\CII] 158 \um\ emission lines and underlying dust continua.
The high angular resolution of these NIRCam and ALMA observations enables spatially resolved analyses at the rest-frame UV, optical, and FIR wavelengths of B14-65666, a bright major merger in the EoR.

B14-65666 consists of two galaxy components, E and W, and diffuse emission surrounding the galaxy E.
The galaxy E is red (UV spectral slope \(\betaUV = -1.2\)) and has a compact core (E-core) with an effective radius of \(r_\text{e} < 0.016\) arcsec (\(85\) pc) at the rest-frame \(1400\) \AA\@.
The galaxy W is blue (\(\betaUV = -2.2\)) and elongated to \(\simeq 0.3\) arcsec (\(1.5\) kpc) with clumpy morphology in the rest-frame UV wavelengths.
Surprisingly, in contrast to their colors, the blue galaxy W is brighter than the red galaxy E in the dust continua at \(88\) and \(158\) \um.
The core of the galaxy E (E-core) is surrounded by diffuse extended emission, which is only detected in the long NIRCam wavelength channels and is also bright in [\OIII] 88 \um.
We assume that this extended emission is likely to be the tidal tails created by the gravitational interactions.
Both of the galaxies E and W show the F444W flux excesses over the F356W fluxes (\(0.84^{+0.09}_{-0.10}\) and \(0.53^{+0.09}_{-0.10}\) mag, respectively), indicating the strong [\OIII]\lamlam4960,5008 emission lines.
The [\OIII] line map estimated from the F356W\(-\)F444W color traces strong [\OIII] emission at the E-core and W-core, with the rest-frame equivalent widths of \(\text{EW}_0 \simeq 2000\) and \(750\) \AA, respectively.
These results show that the major merger has disturbed the morphology of the galaxies and induced the nuclear dusty starbursts.

The SED fitting shows that B14-65666 has a total stellar mass of \(\log M_{*}/\Msun = 9.8\pm0.2\) and \(\SFR_\text{10Myr}\) of \(\simeq 230^{+70}_{-60}\) \Moyr.
These merger-induced starbursts would build young stellar masses comparable to or heavier than the underlying old stellar components.
The stellar-mass ratio of the galaxies E and W spans from 3:1 to 2:1, which confirms the fact that B14-65666 is a major merger in terms of the stellar-mass ratio.
The galaxy E experiences young starbursts with \(\SFR_\text{10Myr} = 180^{+50}_{-40}\) \Moyr\ at ages of \( < 53\) Myr, supported by the high optical [\OIII] emission.
The red color of the galaxy E is explained by the strong dust attenuation of \(A_\text{V} = 1.2\pm0.1\) mag, in opposite to the galaxy W, which has blue color and \(A_\text{V} = 0.3\pm0.1\) mag.
Including ALMA data, the SED fitting explains the red color and faint dust continua of the galaxy E by the scenario that the galaxy E exhibits higher dust temperature than the galaxy W.

As combinations of optical and FIR data, we investigate the IRX-\betaUV\ diagram and the optical-to-FIR [\OIII] line ratios.
The entire system is consistent with the \betaUVO-modified Meurer relation, supporting the Calzetti attenuation law.
The galaxies E and W, however, deviate from the relation; this deviation can be explained by different stellar and dust properties: variations of the dust attenuation law, patchy stellar-and-dust geometry, different dust temperature, and possible effects of the tails.
From the estimated [\OIII]\lamlam4960,5008 map and the [\OIII] 88 \um\ moment-0 maps, we draw the optical-to-FIR [\OIII] line ratio map.
This map illustrates gradual spatial variation in the [\OIII] line ratio, implying spatial variation in the electron temperature and electron density.
From the [\OIII] line ratios, the gas-phase metallicity of the entire galaxy is estimated to be \(\Zgas\sim 0.2\nn0.4\) \Zsun\ by photoionization models under the assumption of the electron density of \(\sim200\nn400\) \cmmm.
The E-core shows higher [\OIII] line ratio than the W-core, indicating lower gas-phase metallicity in the E-core.

In summary, our results support a scenario where B14-65666 is a major merger that gravitationally disturbs the morphology and experiences nuclear dusty starbursts likely triggered by less-enriched gas inflows, after the first (or second) passage.
Although the galaxies E and W have similar UV magnitude, they exhibit different stellar, gaseous, and dust properties and morphology.
The starbursts have indeed enhanced the star-formation activity compared with main-sequence star-forming galaxies at the same redshift.
B14-65666 is an important object that provides us with opportunities to investigate complex stellar buildup processes during major mergers, which are abundant at high redshifts, during the EoR.

\begin{acknowledgments}
We thank Rebecca Bowler for helpful discussions on B14-65666.
We wish to thank the anonymous referee for valuable comments to improve our manuscript.
This research is supported by NAOJ ALMA Scientific Research Grant number 2020-16B.
This paper is a part of the outcome of research performed under a Waseda University Grant for Special Research Projects (Project number: 2024C-478).
JAM, ACG, and LC acknowledge support by grant PIB2021-127718NB-100 from the Spanish Ministry of Science and Innovation/State Agency of Research MCIN/AEI/10.13039/501100011033 and by ``ERDF A way of making Europe''.
TH was supported by Leading Initiative for Excellent Young Researchers, MEXT, Japan (HJH02007) and by JSPS KAKENHI Grant Number 22H01258.
KM, YN, and YT acknowledge financial support from JSPS through KAKENHI grant Number 20K14516, 23KJ0728, and 22H04939, respectively.
CBP acknowledges support by grant CM21\_CAB\_M2\_01 from the Program ``Garant\'{\i}a Juven\'{\i}l'' from the ``Comunidad de Madrid'' 2021.
MPS acknowledges support from grant RYC2021-033094-I funded by MCIN/AEI/10.13039/501100011033 and the European Union NextGenerationEU/PRTR\@.
The JWST data presented in this paper were obtained from the Mikulski Archive for Space Telescopes (MAST) at the Space Telescope Science Institute.
The specific observations analyzed can be accessed via \dataset[https://doi.org/10.17909/zps5-eq82]{https://doi.org/10.17909/zps5-eq82}.
STScI is operated by the Association of Universities for Research in Astronomy, Inc., under NASA contract NAS5–26555.
Support to MAST for these data is provided by the NASA Office of Space Science via grant NAG5–7584 and by other grants and contracts.
This paper makes use of the following ALMA data: ADS/JAO.ALMA\#2016.1.00954.S, ADS/JAO.ALMA\#2017.1.00190.S, and ADS/JAO.ALMA\#2019.1.01491.S.
ALMA is a partnership of ESO (representing its member states), NSF (USA), and NINS (Japan), together with NRC (Canada), MOST and ASIAA (Taiwan), and KASI (Republic of Korea), in cooperation with the Republic of Chile.
The Joint ALMA Observatory is operated by ESO, AUI/NRAO, and NAOJ\@.
This work has made use of data from the European Space Agency (ESA) mission \textit{Gaia} (\url{https://www.cosmos.esa.int/gaia}), processed by the \textit{Gaia} Data Processing and Analysis Consortium (DPAC, \url{https://www.cosmos.esa.int/web/gaia/dpac/consortium}).
Funding for the DPAC has been provided by national institutions, in particular the institutions participating in the \textit{Gaia} Multilateral Agreement.
The \textit{Gaia} data are retrieved from the JVO portal (\url{http://jvo.nao.ac.jp/portal}) operated by the NAOJ\@.
This research has made use of NASA’s Astrophysics Data System.
\end{acknowledgments}

\software{NumPy \citep{Harris.C:2020a}, SciPy \citep{Virtanen.P:2020a}, IPython \citep{Perez.F:2007a}, Matplotlib \citep{Hunter.J:2007a}, Astropy \citep{Astropy-Collaboration:2013a, Astropy-Collaboration:2018a, Astropy-Collaboration:2022a}}
% , photutils \citep{Bradley.L:2023a}, reproject, Cloudy \citep{Ferland.G:2017a}, Bagpipes \citep{Carnall.A:2018a}, PyPHER \citep{Boucaud.A:2016a}, CASA \citep{CASA-Team:2022a}, PyNeb \citep{Luridiana.V:2015a}

\vspace{2cm}
\bibliographystyle{aasjournal}
\bibliography{$HOME/Documents/bibtex/reference} %$

\appendix
\restartappendixnumbering
\renewcommand{\theHfigure}{\thesection\arabic{figure}}

\section{Supplements for the SED fitting}
\label{sec:A1}

\begin{figure*}[t]
    \epsscale{1.2}
    \plotone{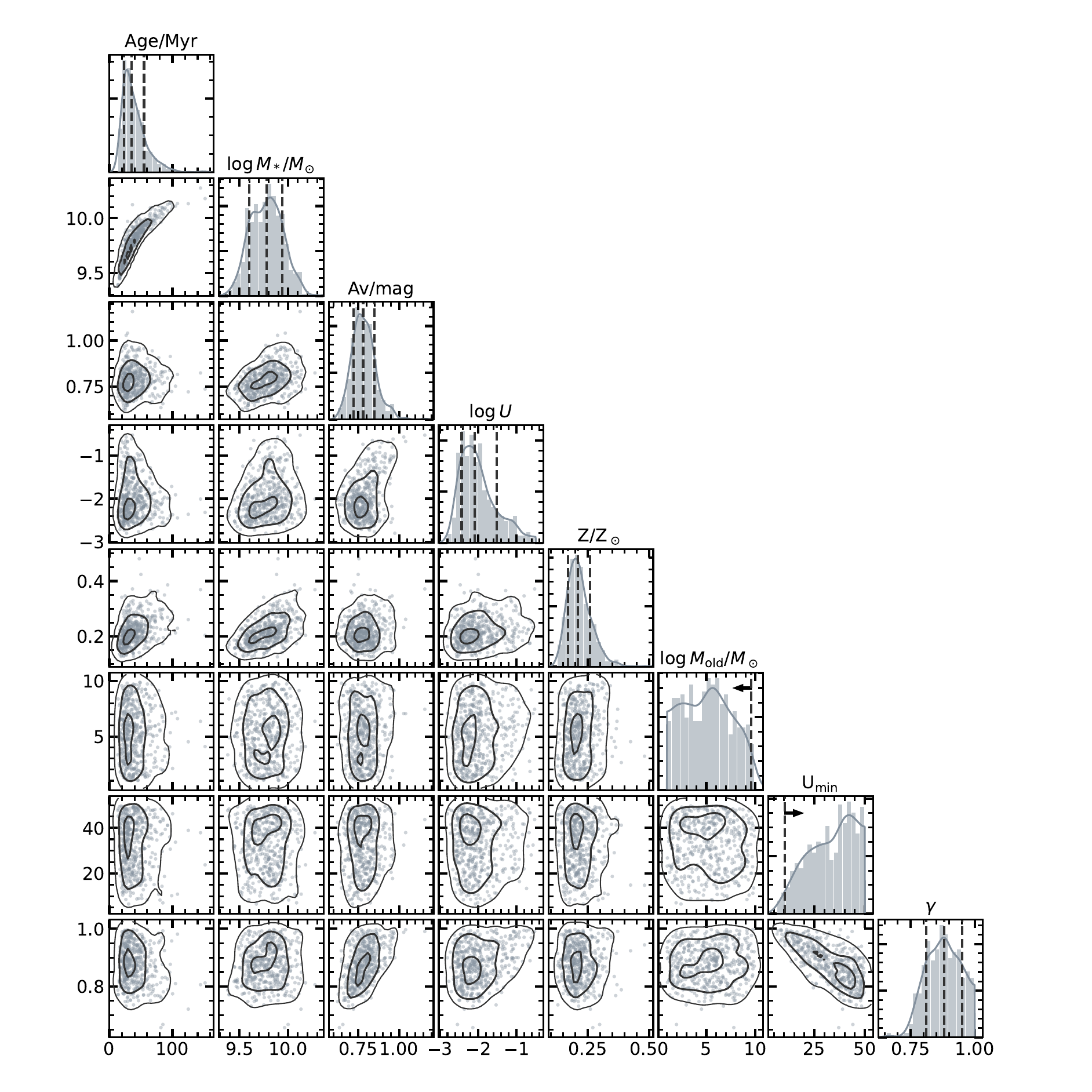}
    \caption{Corner plot for a result of the SED fitting to the total photometry including the ALMA data.
      The gray histograms show the posterior distribution of the fitting parameters and the gray curves show the kernel density estimations KDE of them.
      The vertical three black dashed lines indicate the 16, 50, and 84 percentiles of the posterior distribution of parameters, while the single dashed lines (with the arrows) indicate the \(2\sigma\) lower/upper limits of parameters (i.e., 2/98 percentiles) for posterior distributions that hit to parameter limits and peak close to them.
      The scatter plots show correlations in the posterior distribution between two parameters.
      The black contours show the \(0.2\sigma\), \(1\sigma\), and \(2\sigma\) uncertainty regions computed with KDE.
    }\label{fig:corner}
\end{figure*}

Table \ref{tb:range} shows ranges of the free parameters in the SED fitting, explained in Sec \ref{sec:sed-fitting}.
\texttt{Bagpipes} receives these ranges as limits of the uniform prior distributions and performs the nested sampling to generate the posterior distribution of the parameters.
Figure \ref{fig:corner} shows a corner plot of an obtained posterior distribution in fitting to the total photometry including the ALMA data.

Table \ref{tb:sedsfh} shows results for fitting to the total photometry obtained with different parametric starformation histories: constant (fiducial), delayed \(\tau\), and exponential decreasing histories.
The last row shows the fitting with the constant starformation history but without the old stellar component (i.e., without a parameter \(M_\text{old}\)).
All the starformation histories give the comparable results because B14-65666 exhibits bursty starformation within a short timescale of \(\sim10\) Myr.

\begin{deluxetable}{lcc}[h]
    \tablecaption{Parameter ranges used in the SED fitting.\label{tb:range}}
    \tablewidth{0pt}
    \tablecolumns{2}
    \tablehead{
      \colhead{\hspace{0.5cm}Parameters}\hspace{0.5cm} & \colhead{\hspace{0.5cm}Ranges}\hspace{0.5cm}
    }
    \startdata
    \(t_\text{SF} / \text{Myr}\) & [\(1.0\), \(300.0\)] \\
    \(\mathrm{\log M_{ * } / M_{\sun}}\) & [\(1.0\), \(15.0\)] \\
    \(A_\text{V} / \) mag & [\(0.0\), \(2.0\)] \\
    \(\log U\) & [\(-4.0\), \(-0.5\)] \\
    \(Z_{*} / Z_{\sun}\) & [\(0.01\), \(2.00\)] \\
    \(\mathrm{\log M_\text{ old } / M_{\sun}}\) & [\(1.0\), \(15.0\)] \\
    \(\Umin\) & [\(0.1\), \(50.0\)] \\
    \(\gamma\) & [\(0.001\), \(1.0\)] \\
    \enddata
\end{deluxetable}

%%% Local Variables:
%%% mode: japanese-LaTeX
%%% TeX-master: "../ms"
%%% End:
 % tb:range
\vspace{-1\baselineskip}

\begin{deluxetable*}{lccccccc}[h]
    \tablecaption{Results of the SED fitting with differnt star-formation histories and setups.\label{tb:sedsfh}}
    \tablewidth{0pt}
    \tablecolumns{8}
    \tablehead{
        \colhead{Components} & \colhead{\(\tau\)} & \colhead{\(t_\mathrm{SF}\)} & \colhead{\(\log{M_*}\)} & \colhead{\(A_\mathrm{V}\)} & \colhead{\(\log U\)} & \colhead{\(Z_{*}\)} & \colhead{\(\log{M_\mathrm{old}}\)}\\
        \colhead{} & \colhead{[$\mathrm{Myr}$]} & \colhead{[$\mathrm{Myr}$]} & \colhead{[$\mathrm{\log M_{\odot}}$]} & \colhead{[$\mathrm{mag}$$\mathrm{\left( \mathrm{AB} \right)}$]} & \colhead{} & \colhead{[$\mathrm{Z_{\odot}}$]} & \colhead{[$\mathrm{\log M_{\odot}}$]}
    }
    \startdata
    Constant & - & \(24^{+21}_{-9}\) & \(9.75^{+0.18}_{-0.20}\) & \(0.87^{+0.13}_{-0.11}\) & \(-2.07^{+0.50}_{-0.35}\) & \(0.26^{+0.10}_{-0.07}\) & \(<9.65\)\\
    Delayed \(\tau\) & \(185^{+73}_{-93}\) & \(29^{+29}_{-14}\) & \(9.62^{+0.23}_{-0.19}\) & \(0.88^{+0.14}_{-0.12}\) & \(-2.08^{+0.57}_{-0.34}\) & \(0.28^{+0.10}_{-0.10}\) & \(<9.61\)\\
    Exponential & \(191^{+72}_{-93}\) & \(19^{+13}_{-8}\) & \(9.69^{+0.17}_{-0.23}\) & \(0.89^{+0.13}_{-0.12}\) & \(-2.04^{+0.46}_{-0.38}\) & \(0.27^{+0.09}_{-0.09}\) & \(<9.42\)\\
    Constant (without old) & - & \(15^{+17}_{-9}\) & \(9.60^{+0.25}_{-0.29}\) & \(0.90^{+0.17}_{-0.16}\) & \(-2.07^{+0.65}_{-0.36}\) & \(0.28^{+0.12}_{-0.11}\) & -\\
    \enddata
    \tablecomments{The results of the ``Constant'' starformation history (top) is the same as the ``Total'' with only NIRCam data in Table \ref{tb:sed}. The slight differences between them originate from the numerical uncertainties in the Monte Carlo computations.}
\end{deluxetable*}
 % tb:sedsfh
\vspace{-1\baselineskip}

\end{document}